\documentclass[11pt,english]{article}

\usepackage[latin9]{inputenc}
\usepackage{geometry}
\usepackage{amsmath}
\usepackage{amssymb}
\usepackage{graphicx}
\usepackage{setspace}
\makeatletter

\providecommand{\tabularnewline}{\\}

\newenvironment{lyxlist}[1]
{\begin{list}{}
{\settowidth{\labelwidth}{#1}
 \setlength{\leftmargin}{\labelwidth}
 \addtolength{\leftmargin}{\labelsep}
 }}
{\end{list}}

\@ifundefined{date}{}{\date{}}
\usepackage{array}
\usepackage{mathrsfs}
\usepackage{color}

\makeatother

\usepackage{babel}
\begin{document}

\title{\textbf{Heavy-Tailed Loss Frequencies from}\\
\textbf{ Mixtures of Negative Binomial and Poisson Counts}}

\author{Jiansheng Dai,\thanks{WizardQuant Investment Management; email: daiball@yeah.net.}
\enskip{}Ziheng Huang,\thanks{Sixie Capital Management; email: huangziheng1996@126.com.}
\enskip{}Michael R. Powers,\thanks{Corresponding author; B306 Lihua Building, Department of Finance,
School of Economics and Management, and Schwarzman College, Tsinghua
University, Beijing, China 100084; email: powers@sem.tsinghua.edu.cn.} \enskip{}and Jiaxin Xu\thanks{Department of Finance, School of Economics and Management, Tsinghua
University; email: xujx.18@sem.tsinghua.edu.cn.}}

\date{November 9, 2022}
\maketitle
\begin{abstract}
\begin{singlespace}
\noindent Heavy-tailed random variables have been used in insurance
research to model both loss frequencies and loss severities, with
substantially more emphasis on the latter. In the present work, we
take a step toward addressing this imbalance by exploring the class
of heavy-tailed frequency models formed by continuous mixtures of
Negative Binomial and Poisson random variables. We begin by defining
the concept of a calibrative family of mixing distributions (each
member of which is identifiable from its associated Negative Binomial
mixture), and show how to construct such families from only a single
member. We then introduce a new heavy-tailed frequency model \textendash{}
the two-parameter ZY distribution \textendash{} as a generalization
of both the one-parameter Zeta and Yule distributions, and construct
calibrative families for both the new distribution and the heavy-tailed
two-parameter Waring distribution. Finally, we pursue natural extensions
of both the ZY and Waring families to a unifying, four-parameter heavy-tailed
model, providing the foundation for a novel loss-frequency modeling
approach to complement conventional GLM analyses. This approach is
illustrated by application to a classic set of Swedish commercial
motor-vehicle insurance loss data.\medskip{}

\noindent \textbf{Keywords:} Loss frequency; heavy tail; continuous
mixture; identifiability; calibrative family; Negative Binomial; Poisson.
\end{singlespace}
\end{abstract}

\section{Introduction}

\noindent Let $\mathcal{H}_{\textrm{D}}$ denote the family of heavy-tailed
discrete random variables, $X\in\left\{ 0,1,2,\ldots\right\} $, characterized
by probability mass function (PMF) $f_{X}\left(x\right)>0$, for which
$E_{X}\left[X^{\delta}\right]=\underset{n\rightarrow\infty}{\lim}{\textstyle \sum_{x=0}^{n}}{\displaystyle x^{\delta}f_{X}\left(x\right)}=\infty$
for some real-valued $\delta\in\left(0,\infty\right)$, and let $\mathcal{H}_{\textrm{C}}$
denote the corresponding family of heavy-tailed continuous random
variables, $Y\in\left(0,\infty\right)$, with probability density
function (PDF) $f_{Y}\left(y\right)>0$, such that $E_{Y}\left[Y^{\delta}\right]=\underset{t\rightarrow\infty}{\lim}{\textstyle \int_{0}^{t}}{\displaystyle y^{\delta}f_{Y}\left(y\right)dy}=\infty$
for some positive real $\delta$. Although models from both families
have been employed in the insurance literature, substantially more
attention has been paid to the latter than the former. This difference
may be entirely warranted, simply because individual and total loss
amounts (severities and total losses, respectively) are more likely
to be characterized by heavy tails than are loss counts (frequencies).
However, another consideration \textendash{} which we believe explains
at least part of the difference \textendash{} is that the distributions
in $\mathcal{H}_{\textrm{C}}$ are more familiar and/or more tractable
than those in $\mathcal{H}_{\textrm{D}}$. A principal aim of the
present work is to draw greater attention to heavy-tailed frequency
distributions by exploring the class of heavy-tailed frequency models
formed by continuous mixtures of Negative Binomial (and therefore
Poisson) random variables.

Two of the simplest and best-known sub-families of $\mathcal{H}_{\textrm{D}}$
are:
\begin{lyxlist}{00.00.0000}
\item [{\qquad{}(1)}] \noindent $X|s\sim\textrm{Zeta\ensuremath{\left(s\right)}}$,
with
\begin{equation}
f_{X|s}^{\left(\textrm{Z}\right)}\left(x\right)=\dfrac{\left(x+1\right)^{-s}}{\zeta\left(s\right)},
\end{equation}
for $s\in\left(1,\infty\right)$, where $\zeta\left(s\right)={\textstyle \sum_{x=0}^{\infty}}{\displaystyle \left(x+1\right)^{-s}}$
denotes the Riemann zeta function; and
\item [{\qquad{}(2)}] \noindent $X|b\sim\textrm{Yule}\left(b\right)$,
with
\begin{equation}
f_{X|b}^{\left(\textrm{Y}\right)}\left(x\right)=b\textrm{B}\left(x+1,b+1\right),
\end{equation}
for $b\in\left(0,\infty\right)$, where $\textrm{B}\left(u,w\right)=\tfrac{\Gamma\left(u\right)\Gamma\left(w\right)}{\Gamma\left(u+w\right)}$
denotes the beta function.\footnote{The $\textrm{Zeta}\left(s\right)$ and $\textrm{Yule}\left(b\right)$
distributions often are defined on the sample space $x\in\left\{ 1,2,3,\ldots\right\} $
rather than $x\in\left\{ 0,1,2,\ldots\right\} $. However, we work
with the latter characterization both because it matches the sample
space of the $\textrm{Poisson}\left(\lambda\right)$ distribution
and because it is the more commonly used formulation in insurance
applications.} 
\end{lyxlist}
\noindent Both of these distributions have been proposed to model
loss frequencies,\footnote{See, for example, Doray and Luong (1995) for Zeta applications and
Irwin (1968) for Yule applications (as a special case of the Generalized
Waring distribution).} and they possess comparable properties, including asymptotically
equivalent tails when $b=s-1$. Nevertheless, the similarities and
differences of the two mathematical models have not been analyzed
closely in the literature. In particular, although it is well known
that the Yule distribution (as a special case of the three-parameter
Generalized Waring distribution) can be expressed as a continuous
mixture of Negative Binomial random variables (as shown in Subsection
3.1 below), no comparable result existed for the Zeta distribution
until very recently (see Dai, Huang, Powers, and Xu, 2021). Therefore,
these two types of random variables afford a natural starting point
for the present study.

Following a brief overview of the use of heavy-tailed random variables
in insurance in Section 2, we explore the formation of heavy-tailed
frequencies from continuous mixtures of Negative Binomial and Poisson
random variables. In Section 3, we define the concept of a calibrative
family of mixing distributions (each member of which is identifiable
from its associated Negative Binomial mixture), and show how to construct
such families from only a single member. Then, in Section 4, we introduce
a new heavy-tailed frequency model \textendash{} the two-parameter
ZY distribution \textendash{} as a generalization of both the one-parameter
Zeta and Yule distributions, and construct calibrative families for
both the new distribution and the heavy-tailed two-parameter Waring
distribution, investigating their similarities and differences. Finally,
we pursue natural extensions of both the ZY and Waring families to
a unifying, four-parameter heavy-tailed model, providing the foundation
for a novel loss-frequency modeling approach to complement conventional
GLM analyses. This approach is illustrated by application to a set
of historical Swedish commercial motor-vehicle insurance loss data.

\section{Heavy-Tailed Random Variables in Insurance}

\noindent An obvious place to look for applications of heavy-tailed
loss models is among the damages caused by nature's most destructive
forces: wind, storms, tidal waves, and earthquakes. Indeed, it is
widely recognized that such physical forces exhibit magnitudes that
are well modeled by power laws; and the values of real property \textendash{}
houses, buildings, factories, etc. \textendash{} also tend to follow
such distributions (in much the same way as personal income and wealth).
(See, e.g., Newman, 2005.)

Interestingly, however, heavy-tailed probability models rarely play
an explicit role in contemporary property-catastrophe loss forecasting.
This is not because they are inconsistent with historical data, but
rather because today's insurance industry favors methods relying heavily
on engineering-based simulation, in which the numbers of properties
damaged (frequencies), and their respective insured values, are generated
implicitly through random scenarios. The sizes of individual losses
(severities), then are drawn from (bounded) Beta distributions whose
scale factors (upper bounds) are given by the respective insured values.
Finally, the resulting total-loss random variables \textendash{} although
likely to be heavy-tailed in theory \textendash{} are characterized
by (bounded) empirical distributions.

As a consequence of such practices, the most common applications of
heavy-tailed models do not involve physically forceful loss processes,
but rather the heterogeneity of frequencies and severities associated
with more mundane perils. This heterogeneity can arise from any of
a number of sources, running the gamut from simple to complex. At
the former end of the spectrum, it is easy to envision a commercial
insurance company whose overall loss frequency consists of a mixture
of individual policyholder frequencies, the means of which follow
a power law. At the latter end, one might imagine an insurance company
that, because of underwriting problems, begins to cover policyholders
whose expected frequencies and/or severities follow a heavy-tailed
distribution.

Instructively, even the simplest frequency or severity models can
be converted easily into heavy-tailed random variables by an elementary
transformation attributable to risk heterogeneity.

For example, suppose a commercial policyholder\textquoteright s medical-expense
loss frequency is given by a\linebreak{}
$\textrm{Geometric}\left(p\right)$ random variable; that is, $X\mid p\sim f_{X\mid p}^{\left(\textrm{G}\right)}\left(x\right)=\left(1-p\right)p^{x},\:x\in\left\{ 0,1,2,\ldots\right\} $.\footnote{We parameterize the $\textrm{Negative Binomial}\left(r,p\right)$
distribution \textendash{} of which $\textrm{Geometric}\left(p\right)$
is a special case with $r=1$ \textendash{} using the symbol $p$
to represent the probability of a ``failure'' prior to the $r^{\textrm{th}}$
``success'' (as opposed to the probability of a success, as is more
common). This is because the present parameterization facilitates
certain mathematical derivations (see, e.g., Subsection A.2 of the
Appendix) by permitting the product $\left(1-p\right)p^{x}$ to be
broken into two additive components, $p^{x}-p^{x+1}$ (which cannot
be done with the alternative product, $p\left(1-p\right)^{x}$). An
additional advantage of the less-common parameterization is that both
the mean and variance of the $\textrm{Negative Binomial}\left(r,p\right)$
random variable are increasing functions of $p$, just as the mean
and variance of the $\textrm{Poisson}\left(\lambda\right)$ random
variable are increasing functions of $\lambda$.} Suppose further that the mean frequency, $\mu=E_{X|p}\left[X\right]=\tfrac{p}{1-p}$,
varies with the policyholder's number of employees, and follows a
$\textrm{Pareto 2}\left(\alpha=1,\theta=1\right)$ distribution with
$f_{\mu|\alpha=1,\theta=1}^{\left(\textrm{P}2\right)}\left(\mu\right)=\tfrac{1}{\left(\mu+1\right)^{2}},\:\mu\in\left(0,\infty\right)$.
In that case, $p|a,b=1\sim\textrm{Beta}\left(a,b=1\right)\equiv\textrm{Uniform}\left(0,1\right)$,
and the unconditional medical-expense frequency, $X$, is a $\textrm{Yule}\left(b=1\right)$
random variable with $f_{X|b=1}^{\left(\textrm{Y}\right)}\left(x\right)=\tfrac{1}{\left(x+1\right)\left(x+2\right)}$.

Alternatively, consider a personal-lines policyholder whose liability
loss severity is given by an\linebreak{}
$\textrm{Exponential}\left(\lambda\right)$ random variable, such
that $Y\mid\lambda\sim f_{Y\mid\lambda}^{\left(\textrm{E}\right)}\left(y\right)=\lambda e^{-\lambda y},\:y\in\left(0,\infty\right)$.
Now assume that, because of classification errors and/or information
deficiencies, the mean severity, $\mu=E_{Y|\lambda}\left[Y\right]=\tfrac{1}{\lambda}$,
follows an $\textrm{Inverse Exponential}\left(\theta\right)$ distribution
with $f_{\mu|\theta}^{\left(\textrm{IE}\right)}\left(\mu\right)=\tfrac{\theta}{\mu^{2}}e^{-\theta/\mu},\:\mu\in\left(0,\infty\right)$.
In this case, $\lambda|\theta\sim\textrm{Exponential}\left(\theta\right)$,
and the unconditional liability severity, $Y$, is a $\textrm{Pareto 2}\left(\alpha=1,\theta\right)$
random variable with $f_{Y|\alpha=1,\theta}^{\left(\textrm{P}2\right)}\left(y\right)=\tfrac{\theta}{\left(y+\theta\right)^{2}}$.

As noted in the Introduction, heavy-tailed frequency models are much
less commonly used in actual practice than heavy-tailed severity models.
This does not necessarily mean that heavy-tailed frequencies are rare;
in fact, we would argue that, because of heterogeneous risk levels,
heavy-tailed frequencies are quite common. Rather, the dearth of such
models is likely explained \textendash{} at least in part \textendash{}
by a lack of familiar and tractable heavy-tailed discrete distributions,
and a concomitant shortage of related mathematical analysis. (See,
e.g., Karlis, 2005.) The present work is intended as a step toward
addressing these issues.

\section{Heavy-Tailed Frequency Models}

\subsection{Mixtures of Negative Binomial Counts}

\noindent We begin by focusing on members of $\mathcal{H}_{\textrm{D}}$
that are formed as mixtures of $\textrm{Negative Binomial}\left(r,p\right)$
random variables by treating the $r$ parameter as fixed, and the
$p$ parameter as a continuous random variable, $p|\boldsymbol{\eta}\sim f_{p|\boldsymbol{\eta}}\left(p\right)>0,\:p\in\left(0,1\right)$,
where $\boldsymbol{\eta}$ denotes a vector of distributional parameters.
To that end, let $X\mid r,p\sim\textrm{Negative Binomial}\left(r,p\right)$
with $f_{X\mid r,p}^{\left(\textrm{NB}\right)}\left(x\right)=\tfrac{\Gamma\left(x+r\right)}{\Gamma\left(r\right)\Gamma\left(x+1\right)}\left(1-p\right)^{r}p^{x},\:x\in\left\{ 0,1,2,\ldots\right\} $,
for $r\in\left(0,\infty\right)$ and $p\in\left(0,1\right)$, and
consider the family of mixing PDFs, $\mathcal{M}_{\textrm{NB}}^{\mathcal{H}}$,
such that $f_{p|\boldsymbol{\eta}}\left(p\right)\in\mathcal{M}_{\textrm{NB}}^{\mathcal{H}}\Longrightarrow$
\begin{equation}
X|r,\boldsymbol{\eta}\sim f_{X|r,\boldsymbol{\eta}}\left(x\right)={\textstyle \int_{0}^{1}}\dfrac{\Gamma\left(x+r\right)}{\Gamma\left(r\right)\Gamma\left(x+1\right)}\left(1-p\right)^{r}p^{x}f_{p|\boldsymbol{\eta}}\left(p\right)dp
\end{equation}
is a member of $\mathcal{H}_{\textrm{D}}$.

For simplicity, we will limit membership in $\mathcal{M}_{\textrm{NB}}^{\mathcal{H}}$
to differentiable PDFs that do not oscillate asymptotically as $p$
approaches either 0 or 1 (i.e., $f_{p|\boldsymbol{\eta}}^{\prime}\left(p\right)$
is well defined everywhere on $\left(0,1\right)$, and changes sign
only a finite number of times on this interval). The following result
provides necessary and sufficient conditions for a mixing PDF to be
a member of $\mathcal{M}_{\textrm{NB}}^{\mathcal{H}}$.\medskip{}

\noindent \textbf{Theorem 1:} The following three statements are equivalent:

\noindent (a) $f_{p|\boldsymbol{\eta}}\left(p\right)\in\mathcal{M}_{\textrm{NB}}^{\mathcal{H}}$;

\noindent (b) $\underset{p\rightarrow1^{-}}{\lim}f_{p|\boldsymbol{\eta}}\left(p\right)\left(1-p\right)^{-\delta+1}=L>0$
for some $\delta\in\left(0,\infty\right)$; and

\noindent (c) $\underset{p\rightarrow1^{-}}{\lim}\dfrac{\ln\left(f_{p|\boldsymbol{\eta}}\left(p\right)\right)}{\ln\left(1-p\right)}<\infty$.\medskip{}

\noindent \textbf{Proof:} See Subsection A.1 of the Appendix.\medskip{}

Statement (b) of the above theorem means that $f_{p|\boldsymbol{\eta}}\left(p\right)$
follows an inverse power law in the limit as $p\rightarrow1^{-}$,
which is consistent with the restriction that $f_{p|\boldsymbol{\eta}}^{\prime}\left(p\right)$
changes sign only a finite number of times on $\left(0,1\right)$.
If we initially had permitted asymptotically oscillating functional
forms for $f_{p|\boldsymbol{\eta}}\left(p\right)$, then condition
(b) would represent a narrowing of the set of possible PDFs. The value
of statement (c) comes from its parsimony; that is, the condition
can be expressed without reference to any specific parameter $\delta\in\left(0,\infty\right)$.

One well-studied member of $\mathcal{M}_{\textrm{NB}}^{\mathcal{H}}$
is the $\textrm{Generalized Beta 1}\left(a,b,c\right)$ PDF, $f_{p|a,b,c}^{\left(\textrm{GB}1\right)}\left(p\right)=$\linebreak{}
$\tfrac{c}{\textrm{B}\left(a,b\right)}p^{ca-1}\left(1-p^{c}\right)^{b-1}$,
for $a,b,c\in\left(0,\infty\right)$. This distribution, defined by
McDonald (1984) with an arbitrary positive scale factor, contains
$\textrm{Beta}\left(a,b\right)\equiv\textrm{Generalized Beta 1}\left(a,b,c=1\right)$
and $\textrm{Kumaraswamy}\left(b,c\right)\equiv\textrm{Generalized Beta 1}\left(a=1,b,c\right)$
as special cases. Although we are not aware of any comprehensive investigation
of the general distribution
\begin{equation}
X|r,a,b,c\sim f_{X|r,a,b,c}\left(x\right)={\displaystyle \int_{0}^{1}}f_{X\mid r,p}^{\left(\textrm{NB}\right)}\left(x\right)f_{p|a,b,c}^{\left(\textrm{GB}1\right)}\left(p\right)dp,
\end{equation}
various special cases are well known. These include:
\begin{lyxlist}{00.00.0000}
\item [{\qquad{}(1)}] \noindent $X|r,a,b\sim\textrm{Generalized Waring}\left(r,a,b\right)$,
introduced by Irwin (1968), with
\[
f_{X|r,a,b}^{\left(\textrm{GW}\right)}\left(x\right)={\displaystyle \int_{0}^{1}}f_{X\mid r,p}^{\left(\textrm{NB}\right)}\left(x\right)f_{p|a,b,c=1}^{\left(\textrm{GB}1\right)}\left(p\right)dp
\]
\[
=\dfrac{\textrm{B}\left(x+a,b+r\right)}{\textrm{B}\left(a,b\right)x\textrm{B}\left(x,r\right)};
\]
\item [{\qquad{}(2)}] \noindent $X|a,b\sim\textrm{Waring}\left(a,b\right)\equiv$
$\textrm{Generalized Waring}\left(r=1,a,b\right)$, with
\[
f_{X|a,b}^{\left(\textrm{W}\right)}\left(x\right)={\displaystyle \int_{0}^{1}}f_{X\mid r=1,p}^{\left(\textrm{NB}\right)}\left(x\right)f_{p|a,b,c=1}^{\left(\textrm{GB}1\right)}\left(p\right)dp
\]
\begin{equation}
=\dfrac{\textrm{B}\left(x+a,b+1\right)}{\textrm{B}\left(a,b\right)};
\end{equation}
and
\item [{\qquad{}(3)}] \noindent $X|b\sim\textrm{Yule}\left(b\right)\equiv$
$\textrm{Generalized Waring}\left(r=1,a=1,b\right)$, with
\[
f_{X|b}^{\left(\textrm{Y}\right)}\left(x\right)={\displaystyle \int_{0}^{1}}f_{X\mid r=1,p}^{\left(\textrm{NB}\right)}\left(x\right)f_{p|a=1,b,c=1}^{\left(\textrm{GB}1\right)}\left(p\right)dp
\]
\[
=b\textrm{B}\left(x+1,b+1\right).
\]
\end{lyxlist}
\qquad{}To show that $f_{p|a,b,c}^{\left(\textrm{GB}1\right)}\left(p\right)$
belongs to $\mathcal{M}_{\textrm{NB}}^{\mathcal{H}}$ (and therefore
that $X|r,a,b,c$ of (4) belongs to $\mathcal{H}_{\textrm{D}}$),
we first note that $\underset{p\rightarrow1^{-}}{\lim}\tfrac{1-p^{c}}{1-p}=\underset{p\rightarrow1^{-}}{\lim}cp^{c-1}=c>0$
by a straightforward application of L'Hôpital's rule. It then follows
from condition (a) of Theorem 1 that
\[
\underset{p\rightarrow1^{-}}{\lim}f_{p|a,b,c}^{\left(\textrm{GB}1\right)}\left(p\right)\left(1-p\right)^{-\delta+1}=\underset{p\rightarrow1^{-}}{\lim}\dfrac{c}{\textrm{B}\left(a,b\right)}p^{ca-1}\left(1-p^{c}\right)^{b-1}\left(1-p\right)^{-\delta+1}
\]
\[
=\dfrac{c}{\textrm{B}\left(a,b\right)}\underset{p\rightarrow1^{-}}{\lim}\left(\dfrac{1-p^{c}}{1-p}\right)^{b-1}\left(1-p\right)^{b-1}\left(1-p\right)^{-\delta+1}
\]
\[
=\dfrac{c^{b}}{\textrm{B}\left(a,b\right)}\underset{p\rightarrow1^{-}}{\lim}\left(1-p\right)^{b-\delta},
\]
which is greater than 0 for all $\delta\geq b$. 

Of course, not all PDFs on $\left(0,1\right)$ belong to $\mathcal{M}_{\textrm{NB}}^{\mathcal{H}}$.
An example of a distribution that fails to satisfy the conditions
of the above theorem is the $\textrm{Logit-Normal}\left(\mu,\sigma\right)$
PDF, $f_{p|\mu,\sigma}^{\left(\textrm{LN}\right)}\left(p\right)=\tfrac{1}{\sqrt{2\pi}}\tfrac{1}{p\left(1-p\right)}\exp\left(-\tfrac{\left(\textrm{logit}\left(p\right)-\mu\right)^{2}}{2\sigma^{2}}\right)$,
for $\mu\in\left(-\infty,\infty\right)$ and $\sigma\in\left(0,\infty\right)$.
In this case,
\[
\underset{p\rightarrow1^{-}}{\lim}f_{p|\mu,\sigma}^{\left(\textrm{LN}\right)}\left(p\right)\left(1-p\right)^{-\delta+1}=\underset{p\rightarrow1^{-}}{\lim}\dfrac{1}{\sqrt{2\pi}}\dfrac{1}{p\left(1-p\right)}e^{-\tfrac{\left(\textrm{logit}\left(p\right)-\mu\right)^{2}}{2\sigma^{2}}}\left(1-p\right)^{-\delta+1}
\]
\[
=\dfrac{1}{\sqrt{2\pi}}\underset{p\rightarrow1^{-}}{\lim}\dfrac{1}{\left(1-p\right)^{\delta}}e^{-\tfrac{\left(\textrm{logit}\left(p\right)-\mu\right)^{2}}{2\sigma^{2}}}
\]
\[
=\dfrac{1}{\sqrt{2\pi}}\underset{z\rightarrow\infty}{\lim}\dfrac{1}{\left(e^{z}+1\right)^{-\delta}}e^{-\tfrac{\left(z-\mu\right)^{2}}{2\sigma^{2}}}\textrm{ (for }z=\textrm{logit}\left(p\right))
\]
\[
=\dfrac{1}{\sqrt{2\pi}}\underset{z\rightarrow\infty}{\lim}\dfrac{\left(e^{z}+1\right)^{\delta}}{e^{\tfrac{\left(z-\mu\right)^{2}}{2\sigma^{2}}}}
\]
\[
=\dfrac{1}{\sqrt{2\pi}}\underset{z\rightarrow\infty}{\lim}\dfrac{O\left(e^{\delta z}\right)}{e^{\tfrac{z^{2}}{2\sigma^{2}}}e^{-\tfrac{z\mu}{\sigma^{2}}}e^{\tfrac{\mu^{2}}{2\sigma^{2}}}}
\]
\[
=\dfrac{e^{-\tfrac{\mu^{2}}{2\sigma^{2}}}}{\sqrt{2\pi}}\underset{z\rightarrow\infty}{\lim}\dfrac{O\left(e^{\left(\delta+\tfrac{\mu}{\sigma^{2}}\right)z}\right)}{e^{\tfrac{z^{2}}{2\sigma^{2}}}}
\]
\[
=0
\]
for all $\delta>0$.

\subsection{Identifiability and Calibrative Families}

\noindent When using a Negative Binomial mixture model such as (3),
it may be desirable to know whether or not the mixed random variable
(in this case, $X|r,\boldsymbol{\eta}\sim f_{X|r,\boldsymbol{\eta}}\left(x\right)$)
can be associated with a unique mixing distribution (i.e., $p|\boldsymbol{\eta}\sim f_{p|\boldsymbol{\eta}}\left(p\right)$).
This property, known as identifiability, is necessary if one wishes
to estimate the parameters of the mixing distribution from observations
of the mixed random variable (see, e.g., Xekalaki and Panaretos, 1983).
If the $r$ parameter is fixed, then such mixtures are indeed identifiable
(see Sapatinas, 1995). However, if $r$ is a random variable, then
the mixture
\[
X|\boldsymbol{\xi},\boldsymbol{\eta}\sim f_{X|\boldsymbol{\xi},\boldsymbol{\eta}}\left(x\right)=\int_{0}^{\infty}\int_{0}^{1}f_{X\mid r,p}^{\left(\textrm{NB}\right)}\left(x\right)f_{r,p|\boldsymbol{\xi},\boldsymbol{\eta}}\left(r,p\right)dpdr,
\]
for some joint mixing PDF, $f_{r,p|\boldsymbol{\xi},\boldsymbol{\eta}}\left(r,p\right)$,
is generally not identifiable. For example, if there exists a set
of unique mixing PDFs, $f_{p|r,\boldsymbol{\eta}}\left(p\right)$,
such that $f_{X|\boldsymbol{\eta}}\left(x\right)=f_{X|r,\boldsymbol{\eta}}\left(x\right)=\int_{0}^{1}f_{X\mid r,p}^{\left(\textrm{NB}\right)}\left(x\right)f_{p|r,\boldsymbol{\eta}}\left(p\right)dp$
is invariant over $r\in\mathcal{I}$, for some interval $\mathcal{I}\subset\left(0,\infty\right)$,
then $f_{r,p|\boldsymbol{\xi},\boldsymbol{\eta}}\left(r,p\right)$
cannot be unique because it may by expressed as $f_{r,p|\boldsymbol{\xi},\boldsymbol{\eta}}\left(r,p\right)=f_{r|\boldsymbol{\xi}}\left(r\right)f_{p|r,\boldsymbol{\eta}}\left(p\right)$
for \emph{any} PDF $f_{r|\boldsymbol{\xi}}\left(r\right),\:r\in\mathcal{I}$.

We will call a set of unique mixing PDFs, $f_{p|r,\boldsymbol{\eta}}\left(p\right)$,
with invariant $f_{X|\boldsymbol{\eta}}\left(x\right)=f_{X|r,\boldsymbol{\eta}}\left(x\right)=$\linebreak{}
$\int_{0}^{1}f_{X\mid r,p}^{\left(\textrm{NB}\right)}\left(x\right)f_{p|r,\boldsymbol{\eta}}\left(p\right)dp$
on $r\in\mathcal{I}$, the ``calibrative'' family for $f_{X|\boldsymbol{\eta}}\left(x\right)$
(on $r\in\mathcal{I}$). The following result, which applies to all
types of Negative Binomial mixtures (i.e., not just those belonging
to $\mathcal{H}_{\textrm{D}}$) shows how to construct the calibrative
family for a given $f_{X|\boldsymbol{\eta}}\left(x\right)$, beginning
with only the single member $f_{p|r=1,\boldsymbol{\eta}}\left(p\right)$
(corresponding to the $\textrm{Geometric}\left(p\right)\equiv\textrm{Negative Binomial}\left(r=1,p\right)$
PDF, $f_{X\mid r=1,p}^{\left(\textrm{NB}\right)}\left(x\right)$).
In practical applications, knowing the form of a calibrative family
as a function of $r$ would be useful for testing the fit of $X|\boldsymbol{\eta}\sim f_{X|\boldsymbol{\eta}}\left(x\right)=\int_{0}^{1}f_{X\mid r,p}^{\left(\textrm{NB}\right)}\left(x\right)f_{p|r,\boldsymbol{\eta}}\left(p\right)dp$
for a known, estimated, or hypothesized value of $r$.\medskip{}

\noindent \textbf{Theorem 2:} For a given loss frequency $X|\boldsymbol{\eta}\sim f_{X|\boldsymbol{\eta}}\left(x\right)$,
if there exists a PDF, $f_{p|\boldsymbol{\eta}}\left(p\right)=f_{p|r=1,\boldsymbol{\eta}}\left(p\right)$,
satisfying $f_{X|\boldsymbol{\eta}}\left(x\right)=\int_{0}^{1}f_{X\mid r=1,p}^{\left(\textrm{NB}\right)}\left(x\right)f_{p|\boldsymbol{\eta}}\left(p\right)dp$,
then:

\noindent (1) $f_{p|\boldsymbol{\eta}}\left(p\right)$ is unique;

\noindent (2) for all $r\in\left(1,\infty\right)$, the function
\begin{equation}
f_{p|r>1,\boldsymbol{\eta}}\left(p\right)=\dfrac{\left(r-1\right)}{\left(1-p\right)^{r}}{\displaystyle \int_{p}^{1}}\dfrac{\left(\omega-p\right)^{r-2}\left(1-\omega\right)}{\omega^{r-1}}f_{p|\boldsymbol{\eta}}\left(\omega\right)d\omega
\end{equation}
is the unique PDF satisfying $f_{X|\boldsymbol{\eta}}\left(x\right)=\int_{0}^{1}f_{X\mid r>1,p}^{\left(\textrm{NB}\right)}\left(x\right)f_{p|r>1,\boldsymbol{\eta}}\left(p\right)dp$;
and

\noindent (3) for all $r\in\left(0,1\right)$, the function
\[
f_{p|r<1,\boldsymbol{\eta}}\left(p\right)=\dfrac{1}{\left(1-p\right)^{r}}{\displaystyle \int_{p}^{1}}\dfrac{\left(\omega-p\right)^{r-1}}{\omega^{r-1}}\left[1+\left(r-1\right)\dfrac{\left(1-\omega\right)}{\omega}\right]f_{p|\boldsymbol{\eta}}\left(\omega\right)d\omega
\]
\begin{equation}
-\dfrac{1}{\left(1-p\right)^{r}}{\displaystyle \int_{p}^{1}}\dfrac{\left(\omega-p\right)^{r-1}\left(1-\omega\right)}{\omega^{r-1}}f_{p|\boldsymbol{\eta}}^{\prime}\left(\omega\right)d\omega
\end{equation}
is either the unique PDF or a quasi-PDF (such that $f_{p|r<1,\boldsymbol{\eta}}\left(p\right)<0$
for some $p\in\left(0,1\right)$) satisfying $f_{X|\boldsymbol{\eta}}\left(x\right)=\int_{0}^{1}f_{X\mid r<1,p}^{\left(\textrm{NB}\right)}\left(x\right)f_{p|r<1,\boldsymbol{\eta}}\left(p\right)dp$.\medskip{}

\noindent \textbf{Proof:} See Subsection A.2 of the Appendix.\medskip{}

Naturally, the appearance of quasi-PDFs in part (3) of the above theorem
is somewhat surprising. The result below, which provides both necessary
and sufficient conditions for $f_{p|r<1,\boldsymbol{\eta}}\left(p\right)<0$,
shows that quasi-PDFs arise because of the behavior of the original
mixing PDF, $f_{p|\boldsymbol{\eta}}\left(p\right)=f_{p|r=1,\boldsymbol{\eta}}\left(p\right)$,
for values of $p$ in a neighborhood of 0. Thus, this issue is not
directly related to that of heavy tails, which (as we know from Theorem
1) are associated with the behavior of $f_{p|\boldsymbol{\eta}}\left(p\right)$
for values of $p$ close to 1.\medskip{}

\noindent \textbf{Theorem 3: }The function $f_{p|r<1,\boldsymbol{\eta}}\left(p\right)$
of part (3) of Theorem 2 is characterized by the following two pairs
of necessary and sufficient conditions:

\noindent (1) $f_{p|r<1,\boldsymbol{\eta}}\left(p\right)$ is a quasi-PDF
with $f_{p|r<1,\boldsymbol{\eta}}\left(p\right)\leq\ell<0$ in some
neighborhood of 0 if

\noindent \qquad{}(a) either (i) $\underset{p\rightarrow0^{+}}{\lim}f_{p|\boldsymbol{\eta}}\left(p\right)<\infty$
or (ii) $\underset{p\rightarrow0^{+}}{\lim}f_{p|\boldsymbol{\eta}}\left(p\right)=\infty$
$\cap$ $\underset{p\rightarrow0^{+}}{\lim}\dfrac{-pf_{p|\boldsymbol{\eta}}^{\prime}\left(p\right)}{f_{p|\boldsymbol{\eta}}\left(p\right)}<1-r$;

\noindent and only if

\noindent \qquad{}(b) either (i) $\underset{p\rightarrow0^{+}}{\lim}f_{p|\boldsymbol{\eta}}\left(p\right)<\infty$
or (ii) $\underset{p\rightarrow0^{+}}{\lim}f_{p|\boldsymbol{\eta}}\left(p\right)=\infty$
$\cap$ $\underset{p\rightarrow0^{+}}{\lim}\dfrac{-pf_{p|\boldsymbol{\eta}}^{\prime}\left(p\right)}{f_{p|\boldsymbol{\eta}}\left(p\right)}\leq1-r$.

\noindent (2) $f_{p|r<1,\boldsymbol{\eta}}\left(p\right)$ is the
unique PDF if

\noindent \qquad{}(a) $\underset{p\rightarrow0^{+}}{\lim}f_{p|\boldsymbol{\eta}}\left(p\right)=\infty$
$\cap$ $\underset{p\rightarrow0^{+}}{\lim}\dfrac{-pf_{p|\boldsymbol{\eta}}^{\prime}\left(p\right)}{f_{p|\boldsymbol{\eta}}\left(p\right)}\geq1-r$
$\cap$ $\dfrac{-pf_{p|\boldsymbol{\eta}}^{\prime}\left(p\right)}{f_{p|\boldsymbol{\eta}}\left(p\right)}>1-r-\dfrac{p}{1-p}\textrm{ for all }p\in\left(0,1\right)$,

\noindent and only if

\noindent \qquad{}(b) $\underset{p\rightarrow0^{+}}{\lim}f_{p|\boldsymbol{\eta}}\left(p\right)=\infty$
$\cap$ $\underset{p\rightarrow0^{+}}{\lim}\dfrac{-pf_{p|\boldsymbol{\eta}}^{\prime}\left(p\right)}{f_{p|\boldsymbol{\eta}}\left(p\right)}\geq1-r$.\medskip{}

\noindent \textbf{Proof:} See Subsection A.3 of the Appendix.\medskip{}

It is worth noting that the sufficient condition provided by (1)(a)
of the above theorem can be sharpened by adding constraints on higher-order
derivatives of $f_{p|\boldsymbol{\eta}}\left(p\right)$ that imply
$\underset{p\rightarrow0^{+}}{\lim}f_{p|r<1,\boldsymbol{\eta}}\left(p\right)<0$
when $\underset{p\rightarrow0^{+}}{\lim}f_{p|\boldsymbol{\eta}}\left(p\right)=\infty$
and $\underset{p\rightarrow0^{+}}{\lim}\tfrac{-pf_{p|\boldsymbol{\eta}}^{\prime}\left(p\right)}{f_{p|\boldsymbol{\eta}}\left(p\right)}=1-r$.
For example, applying L'Hôpital's rule to the indeterminate form of
(A11) (in the Appendix) reveals that
\[
\underset{p\rightarrow0^{+}}{\lim}f_{p|\boldsymbol{\eta}}\left(p\right)=\infty\:\cap\:\underset{p\rightarrow0^{+}}{\lim}\dfrac{-pf_{p|\boldsymbol{\eta}}^{\prime}\left(p\right)}{f_{p|\boldsymbol{\eta}}\left(p\right)}=1-r\:\cap\:\underset{p\rightarrow0^{+}}{\lim}\dfrac{-pf_{p|\boldsymbol{\eta}}^{\prime\prime}\left(p\right)}{f_{p|\boldsymbol{\eta}}^{\prime}\left(p\right)}<2-r
\]
also is a sufficient condition for a quasi-PDF with negative values
in a neighborhood of 0. However, the usefulness of such incremental
improvements is limited because it is relatively easy to find mixing
PDFs, $f_{p|\boldsymbol{\eta}}\left(p\right)$, that fail to satisfy
the enhanced conditions but still yield $\underset{p\rightarrow0^{+}}{\lim}f_{p|r<1,\boldsymbol{\eta}}\left(p\right)<0$.
A simple example is given by the $\textrm{Kumaraswamy}\left(b,c\right)$
PDF with $c=r$,
\[
f_{p|b,c=r}^{\left(\textrm{K}\right)}\left(p\right)=rbp^{r-1}\left(1-p^{r}\right)^{b-1},
\]
for which $\underset{p\rightarrow0^{+}}{\lim}f_{p|\boldsymbol{\eta}}\left(p\right)=\infty$,
$\underset{p\rightarrow0^{+}}{\lim}\tfrac{-pf_{p|\boldsymbol{\eta}}^{\prime}\left(p\right)}{f_{p|\boldsymbol{\eta}}\left(p\right)}=1-r$,
and $\underset{p\rightarrow0^{+}}{\lim}\tfrac{-pf_{p|\boldsymbol{\eta}}^{\prime\prime}\left(p\right)}{f\prime_{p|\boldsymbol{\eta}}\left(p\right)}=2-r$,
but
\[
\underset{p\rightarrow0^{+}}{\lim}f_{p|r<1,\boldsymbol{\eta}}\left(p\right)=1+r^{2}b\left(b-1\right)\underset{p\rightarrow0^{+}}{\lim}{\displaystyle \int_{p}^{1}}\dfrac{\left(1-\omega\right)}{\left(1-\omega^{r}\right)\omega}\omega^{2r-1}\left(1-\omega^{r}\right)^{b-1}d\omega
\]
\[
=-\infty
\]
for all $r\in\left(0,1/2\right)$ and $b\in\left(0,1\right)$.

\subsection{Mixtures of Poisson Counts}

\noindent It is well known that any Negative Binomial random variable
can be expressed as a unique continuous mixture of Poisson random
variables. Specifically,
\begin{equation}
f_{X\mid r,p}^{\left(\textrm{NB}\right)}\left(x\right)=\int_{0}^{\infty}f_{X\mid\lambda}^{\left(\textrm{P}\right)}\left(x\right)f_{\lambda|r,\tfrac{1-p}{p}}^{\left(\Gamma\right)}\left(\lambda\right)d\lambda,
\end{equation}
where $f_{X\mid\lambda}^{\left(\textrm{P}\right)}\left(x\right)=\tfrac{e^{-\lambda}\lambda^{x}}{x!},\:x\in\left\{ 0,1,2,\ldots\right\} $
and $f_{\lambda|r,\tfrac{1-p}{p}}^{\left(\Gamma\right)}\left(\lambda\right)=\tfrac{1}{\Gamma\left(r\right)}\left(\tfrac{1-p}{p}\right)^{r}\lambda^{r-1}\exp\left(-\left(\tfrac{1-p}{p}\right)\lambda\right),\:\lambda\in\left(0,\infty\right)$
denote the $\textrm{Poisson}\left(\lambda\right)$ PMF and $\textrm{Gamma}\left(r,\tfrac{1-p}{p}\right)$
PDF, respectively. This fact, in conjunction with Theorem 2, allows
one to show that any Negative Binomial mixture $f_{X|\boldsymbol{\eta}}\left(x\right)=\int_{0}^{1}f_{X\mid r,p}^{\left(\textrm{NB}\right)}\left(x\right)f_{p|r,\boldsymbol{\eta}}\left(p\right)dp$
with calibrative family $f_{p|r,\boldsymbol{\eta}}\left(p\right)$
also can be expressed as a unique Poisson mixture, $f_{X|\boldsymbol{\eta}}\left(x\right)=\int_{0}^{\infty}f_{X\mid\lambda}^{\left(\textrm{P}\right)}\left(x\right)f_{\lambda|\boldsymbol{\eta}}\left(\lambda\right)d\lambda$.
In practical applications, knowing the form of the PDF $f_{\lambda|\boldsymbol{\eta}}\left(\lambda\right)$
would be useful for testing the fit of $X|\boldsymbol{\eta}\sim f_{X|\boldsymbol{\eta}}\left(x\right)=\int_{0}^{\infty}f_{X\mid\lambda}^{\left(\textrm{P}\right)}\left(x\right)f_{\lambda|\boldsymbol{\eta}}\left(\lambda\right)d\lambda$.\medskip{}

\noindent \textbf{Theorem 4:} For a given loss frequency $X|\boldsymbol{\eta}\sim f_{X|\boldsymbol{\eta}}\left(x\right)$,
if there exists a PDF, $f_{p|\boldsymbol{\eta}}\left(p\right)=f_{p|r=1,\boldsymbol{\eta}}\left(p\right)$,
satisfying $f_{X|\boldsymbol{\eta}}\left(x\right)=\int_{0}^{1}f_{X\mid r=1,p}^{\left(\textrm{NB}\right)}\left(x\right)f_{p|\boldsymbol{\eta}}\left(p\right)dp$,
then the function
\[
f_{\lambda|\boldsymbol{\eta}}\left(\lambda\right)={\textstyle \int_{0}^{1}}\left(\dfrac{1-p}{p}\right)\exp\left(-\left(\dfrac{1-p}{p}\right)\lambda\right)f_{p|\boldsymbol{\eta}}\left(p\right)dp
\]
is the unique PDF satisfying $f_{X|\boldsymbol{\eta}}\left(x\right)=\int_{0}^{\infty}f_{X\mid\lambda}^{\left(\textrm{P}\right)}\left(x\right)f_{\lambda|\boldsymbol{\eta}}\left(\lambda\right)d\lambda$.\medskip{}

\noindent \textbf{Proof:} See Subsection A.4 of the Appendix.

\section{The $\mathbf{ZY}\boldsymbol{\left(b,c\right)}$ Distribution}

\subsection{Definition}

\noindent We now introduce the discrete two-parameter ``$\textrm{ZY}\left(b,c\right)$''
distribution, named for the Zeta and Yule families it contains as
special cases. Let $X|b,c\sim\textrm{ZY}\left(b,c\right)$ be defined
by its PMF,
\begin{equation}
f_{X|b,c}^{\left(\textrm{ZY}\right)}\left(x\right)=\dfrac{\textrm{B}\left(\dfrac{x+1}{c},b+1\right)}{\Sigma_{\textrm{B}}\left(\dfrac{1}{c},\dfrac{1}{c},b\right)}
\end{equation}
for $x\in\left\{ 0,1,2,\ldots\right\} $, with $b\in\left(0,\infty\right)$
and $c\in\left(0,\infty\right)$, where
\begin{equation}
\Sigma_{\textrm{B}}\left(\gamma,u,w\right)=\sum_{k=0}^{\infty}\textrm{B}\left(\gamma k+u,w+1\right).
\end{equation}
Given the relatively complex nature of $\Sigma_{\textrm{B}}\left(\dfrac{1}{c},\dfrac{1}{c},b\right)$,
it often is easier to work with the PMF ratio, $\tfrac{f_{X|b,c}^{\left(\textrm{ZY}\right)}\left(x\right)}{f_{X|b,c}^{\left(\textrm{ZY}\right)}\left(x+1\right)}$,
for arbitrary $x\in\left\{ 0,1,2,\ldots\right\} $, rather than the
PMF itself.

From (9) and (10), it is clear that both $f_{X|b,c}^{\left(\textrm{ZY}\right)}\left(x\right)\geq0$
for all $x\in\left\{ 0,1,2,\ldots\right\} $ and ${\textstyle \sum_{x=0}^{\infty}}f_{X|b,c}^{\left(\textrm{ZY}\right)}\left(x\right)=1$.
Furthermore, the PMF is not only well-behaved for all values of its
sample and parameter spaces, but also in the limit as $c\rightarrow0^{+}$.

Taking the relevant limit of the PMF ratio yields
\[
\dfrac{f_{X|b,c=0}^{\left(\textrm{ZY}\right)}\left(x\right)}{f_{X|b,c=0}^{\left(\textrm{ZY}\right)}\left(x+1\right)}=\underset{c\rightarrow0^{+}}{\lim}\dfrac{\Gamma\left(\dfrac{x+1}{c}\right)\Gamma\left(\dfrac{x+2}{c}+b+1\right)}{\Gamma\left(\dfrac{x+1}{c}+b+1\right)\Gamma\left(\dfrac{x+2}{c}\right)}
\]
\[
=\underset{\tau\rightarrow\infty}{\lim}\dfrac{\Gamma\left(\tau\left(x+1\right)\right)\Gamma\left(\tau\left(x+2\right)+b+1\right)}{\Gamma\left(\tau\left(x+1\right)+b+1\right)\Gamma\left(\tau\left(x+2\right)\right)}
\]
\begin{equation}
=\underset{\tau\rightarrow\infty}{\lim}\dfrac{\sqrt{\dfrac{2\pi}{\tau\left(x+1\right)}}\left(\dfrac{\tau\left(x+1\right)}{e}\right)^{\tau\left(x+1\right)}\sqrt{\dfrac{2\pi}{\tau\left(x+2\right)+b+1}}\left(\dfrac{\tau\left(x+2\right)+b+1}{e}\right)^{\tau\left(x+2\right)+b+1}}{\sqrt{\dfrac{2\pi}{\tau\left(x+1\right)+b+1}}\left(\dfrac{\tau\left(x+1\right)+b+1}{e}\right)^{\tau\left(x+1\right)+b+1}\sqrt{\dfrac{2\pi}{\tau\left(x+2\right)}}\left(\dfrac{\tau\left(x+2\right)}{e}\right)^{\tau\left(x+2\right)}},
\end{equation}
where (11) is obtained by replacing all gamma functions with Stirling's
approximation. This expression then simplifies to
\[
\underset{\tau\rightarrow\infty}{\lim}\sqrt{\dfrac{\tau\left(x+2\right)}{\tau\left(x+1\right)}}\sqrt{\dfrac{\tau\left(x+1\right)+b+1}{\tau\left(x+2\right)+b+1}}\left(\dfrac{\tau\left(x+1\right)}{\tau\left(x+1\right)+b+1}\right)^{\tau\left(x+1\right)}
\]
\[
\times\left(\dfrac{\tau\left(x+2\right)+b+1}{\tau\left(x+2\right)}\right)^{\tau\left(x+2\right)}\left(\dfrac{\tau\left(x+2\right)+b+1}{\tau\left(x+1\right)+b+1}\right)^{b+1}
\]
\[
=\underset{\tau\rightarrow\infty}{\lim}\sqrt{\dfrac{\tau\left(x+2\right)}{\tau\left(x+1\right)}}\sqrt{\dfrac{\tau\left(x+1\right)+b+1}{\tau\left(x+2\right)+b+1}}\left(1-\dfrac{b+1}{\tau\left(x+1\right)+b+1}\right)^{\tau\left(x+1\right)+b+1}
\]
\[
\times\left(1+\dfrac{b+1}{\tau\left(x+2\right)}\right)^{\tau\left(x+2\right)}\left(\dfrac{\tau\left(x+2\right)+b+1}{\tau\left(x+1\right)}\right)^{b+1}
\]
\[
=\sqrt{\dfrac{x+2}{x+1}}\sqrt{\dfrac{x+1}{x+2}}e^{-\left(b+1\right)}e^{b+1}\left(\dfrac{x+2}{x+1}\right)^{b+1}
\]
\[
=\left(\dfrac{x+2}{x+1}\right)^{b+1},
\]
which is the corresponding PMF ratio from (1) for $s=b+1$. Thus,
we will say that $X|b\sim\textrm{Zeta}\left(b+1\right)$ constitutes
a special case of $X|b,c\sim\textrm{ZY}\left(b,c\right)$ as $c\rightarrow0^{+}$.

Setting $c=1$ in (9) yields a more immediate result:
\[
\dfrac{f_{X|b,c=1}^{\left(\textrm{ZY}\right)}\left(x\right)}{f_{X|b,c=1}^{\left(\textrm{ZY}\right)}\left(x+1\right)}=\dfrac{\Gamma\left(x+1\right)\Gamma\left(x+b+3\right)}{\Gamma\left(x+b+2\right)\Gamma\left(x+2\right)},
\]
which is the indicated PMF ratio from (2). Consequently, we also can
conclude that $X|b\sim\textrm{Yule\ensuremath{\left(b\right)}}$ forms
a special case of $X|b,c\sim\textrm{ZY}\left(b,c\right)$ for $c=1$.

\subsection{Mixtures of Negative Binomial and Poisson Counts}

\noindent It is instructive to show that $X|b,c\sim\textrm{ZY}\left(b,c\right)$
\textendash{} and therefore $X|b\sim\textrm{Zeta}\left(b+1\right)$
and $X|b\sim\textrm{Yule}\left(b\right)$ \textendash{} can be formed
as unique, continuous mixtures of both Negative Binomial and Poisson
random variables. This is accomplished by first finding a PDF, $f_{p|b,c}\left(p\right)$,
such that $f_{X|b,c}^{\left(\textrm{ZY}\right)}\left(x\right)=\int_{0}^{1}f_{X\mid r=1,p}^{\left(\textrm{NB}\right)}\left(x\right)f_{p|b,c}\left(p\right)dp$,
and then applying the results of Theorems 2 and 4 to identify $f_{p|r,b,c}\left(p\right)$
and $f_{\lambda|b,c}\left(p\right)$, respectively. 

Part (A) of the following result provides what we will call the ``$\textrm{Sigma-B}\left(b,c\right)$''
mixing distribution, $f_{p|b,c}^{\left(\Sigma_{\textrm{B}}\right)}\left(p\right)\sim\Sigma_{\textrm{B}}\left(b,c\right)$,
for $c\in\left(0,\infty\right)$, and part (B) addresses the limiting
case of $c\rightarrow0^{+}$.\medskip{}

\noindent \textbf{Theorem 5:}

\noindent (A) For $X|b,c>0\sim\textrm{ZY}\left(b,c>0\right)$, the
function
\begin{equation}
f_{p|b,c>0}^{\left(\Sigma_{\textrm{B}}\right)}\left(p\right)=\dfrac{c}{\Sigma_{\textrm{B}}\left(\dfrac{1}{c},\dfrac{1}{c},b\right)}\dfrac{\left(1-p^{c}\right)^{b}}{\left(1-p\right)},
\end{equation}
with $\Sigma_{\textrm{B}}\left(\dfrac{1}{c},\dfrac{1}{c},b\right)$
as defined in (17), is the unique PDF satisfying $f_{X|b,c>0}^{\left(\textrm{ZY}\right)}\left(x\right)=\int_{0}^{1}f_{X\mid r=1,p}^{\left(\textrm{NB}\right)}\left(x\right)f_{p|b,c>0}^{\left(\Sigma_{\textrm{B}}\right)}\left(p\right)dp$.

\noindent (B) For $X|b,c\rightarrow0\sim\textrm{ZY}\left(b,c\rightarrow0\right)$,
the function
\begin{equation}
f_{p|b,c\rightarrow0}^{\left(\Sigma_{\textrm{B}}\right)}\left(p\right)=\dfrac{\left(-\ln\left(p\right)\right)^{b}}{\zeta\left(b+1\right)\Gamma\left(b+1\right)\left(1-p\right)}
\end{equation}
is the unique PDF satisfying $f_{X|b,c\rightarrow0}^{\left(\textrm{ZY}\right)}\left(x\right)=\int_{0}^{1}f_{X\mid r=1,p}^{\left(\textrm{NB}\right)}\left(x\right)f_{p|b,c\rightarrow0}^{\left(\Sigma_{\textrm{B}}\right)}\left(p\right)dp$.\medskip{}

\noindent \textbf{Proof:} See Subsection A.5 of the Appendix.\medskip{}

The next result, addressing Negative Binomial mixtures, follows directly
from Theorem 2. Again, part (A) considers the case of $c\in\left(0,\infty\right)$,
and part (B) the limiting case of $c\rightarrow0^{+}$.\medskip{}

\noindent \textbf{Corollary 1:}

\noindent (A) If $X|b,c>0\sim\textrm{ZY}\left(b,c>0\right)$, then:

\noindent (1) for all $r\in\left(1,\infty\right)$, the function
\[
f_{p|r>1,b,c>0}\left(p\right)=\dfrac{c\left(r-1\right)}{\Sigma_{\textrm{B}}\left(\dfrac{1}{c},\dfrac{1}{c},b\right)\left(1-p\right)^{r}}{\displaystyle \int_{p}^{1}}\dfrac{\left(\omega-p\right)^{r-2}\left(1-\omega^{c}\right)^{b}}{\omega^{r-1}}d\omega
\]
is the unique PDF satisfying $f_{X|b,c>0}^{\left(\textrm{ZY}\right)}\left(x\right)=\int_{0}^{1}f_{X\mid r>1,p}^{\left(\textrm{NB}\right)}\left(x\right)f_{p|r>1,b,c>0}\left(p\right)dp$;
and

\noindent (2) for all $r\in\left(0,1\right)$, the function
\[
f_{p|r<1,b,c>0}\left(p\right)=\dfrac{c}{\Sigma_{\textrm{B}}\left(\dfrac{1}{c},\dfrac{1}{c},b\right)\left(1-p\right)^{r}}
\]
\[
\times\left[bc{\displaystyle \int_{p}^{1}}\dfrac{\left(\omega-p\right)^{r-1}\left(1-\omega^{c}\right)^{b-1}}{\omega^{r-c}}d\omega+\left(r-1\right){\displaystyle \int_{p}^{1}}\dfrac{\left(\omega-p\right)^{r-1}\left(1-\omega^{c}\right)^{b}}{\omega^{r}}d\omega\right]
\]
is a quasi-PDF satisfying $f_{X|b,c>0}^{\left(\textrm{ZY}\right)}\left(x\right)=\int_{0}^{1}f_{X\mid r<1,p}^{\left(\textrm{NB}\right)}\left(x\right)f_{p|r<1,b,c>0}\left(p\right)dp$,
with $f_{p|r<1,b,c>0}\left(p\right)\leq\ell<0$ for all $p$ in some
neighborhood of 0.

\noindent (B) If $X|b,c\rightarrow0\sim\textrm{ZY}\left(b,c\rightarrow0\right)$,
then:

\noindent (1) for all $r\in\left(1,\infty\right)$, the function
\[
f_{p|r>1,b,c\rightarrow0}\left(p\right)=\dfrac{\left(r-1\right)}{\zeta\left(b+1\right)\Gamma\left(b+1\right)\left(1-p\right)^{r}}{\displaystyle \int_{p}^{1}}\dfrac{\left(\omega-p\right)^{r-2}\left(-\ln\left(\omega\right)\right)^{b}}{\omega^{r-1}}d\omega
\]
is the unique PDF satisfying $f_{X|b,c\rightarrow0}^{\left(\textrm{ZY}\right)}\left(x\right)=\int_{0}^{1}f_{X\mid r>1,p}^{\left(\textrm{NB}\right)}\left(x\right)f_{p|r>1,b,c\rightarrow0}\left(p\right)dp$;
and

\noindent (2) for all $r\in\left(0,1\right)$, the function
\[
f_{p|r<1,b,c\rightarrow0}\left(p\right)=\dfrac{1}{\zeta\left(b+1\right)\Gamma\left(b+1\right)\left(1-p\right)^{r}}
\]
\[
\times\left[b{\displaystyle \int_{p}^{1}}\dfrac{\left(\omega-p\right)^{r-1}\left(-\ln\left(\omega\right)\right)^{b-1}}{\omega^{r}}d\omega+\left(r-1\right){\displaystyle \int_{p}^{1}}\dfrac{\left(\omega-p\right)^{r-1}\left(-\ln\left(\omega\right)\right)^{b}}{\omega^{r}}d\omega\right]
\]
is a quasi-PDF satisfying $f_{X|b,c\rightarrow0}^{\left(\textrm{ZY}\right)}\left(x\right)=\int_{0}^{1}f_{X\mid r<1,p}^{\left(\textrm{NB}\right)}\left(x\right)f_{p|r<1,b,c\rightarrow0}\left(p\right)dp$,
with $f_{p|r<1,b,c\rightarrow0}\left(p\right)\leq\ell<0$ for all
$p$ in some neighborhood of 0.\medskip{}

\noindent \textbf{Proof:} See Subsection A.6 of the Appendix.\medskip{}

The intuition behind $f_{p|r<1,b,c>0}\left(p\right)<0$ and $f_{p|r<1,b,c\rightarrow0}\left(p\right)<0$
in certain neighborhoods of 0 is fairly straightforward. Essentially,
when $r\in\left(0,1\right)$, the Negative Binomial PMF becomes very
steep for values of $x$ close to (and including) 0 (e.g., $\underset{r}{\sup}\left(\tfrac{f_{X|r,p}^{\left(\textrm{NB}\right)}\left(0\right)}{f_{X|r,p}^{\left(\textrm{NB}\right)}\left(1\right)}\right)=\underset{r\rightarrow0^{+}}{\lim}\tfrac{1}{rp}=\infty$),
and this steepness is aggravated for values of $p$ close to zero.
Consequently, for small values of $x$, it is impossible to construct
the much flatter ZY PMF (for which both $\underset{b}{\inf}\left(\tfrac{f_{X|b,c>0}^{\left(\textrm{ZY}\right)}\left(0\right)}{f_{X|b,c>0}^{\left(\textrm{ZY}\right)}\left(1\right)}\right)$
$=\underset{b\rightarrow0^{+}}{\lim}\tfrac{\Gamma\left(\tfrac{1}{c}\right)\Gamma\left(\tfrac{2}{c}+b+1\right)}{\Gamma\left(\tfrac{1}{c}+b+1\right)\Gamma\left(\tfrac{2}{c}\right)}=2$
and $\underset{b}{\inf}\left(\tfrac{f_{X|b,c\rightarrow0}^{\left(\textrm{ZY}\right)}\left(0\right)}{f_{X|b,c\rightarrow0}^{\left(\textrm{ZY}\right)}\left(1\right)}\right)$
$=\underset{b\rightarrow0^{+}}{\lim}2^{b+1}=2$) as a convex combination
of Negative Binomial PMFs. However, if one can assign negative weight
to those Negative Binomial PMFs for which $p$ is very small, then
the impact of small $r$ can be mitigated by offsetting it with negative
contributions from small $p$.

Although we previously showed that $X|b,c\sim\textrm{ZY}\left(b,c\right)$
is heavy-tailed for both $c\rightarrow0^{+}$ and $c=1$ (by virtue
of the fact that both the Zeta and Yule distributions, respectively,
are heavy-tailed), a direct proof that $\textrm{ZY}\left(b,c\right)\in\mathcal{H}_{\textrm{D}}$
for all $c>0$ was not readily apparent. However, given that $X|b,c>0\sim\textrm{ZY}\left(b,c>0\right)$
can be constructed as mixtures of $\textrm{Negative Binomial}\left(r=1,p\right)$
random variables by part (A) of Theorem 5, it now is possible to demonstrate
the heavy-tailed nature of all ZY random variables for $c>0$ by applying
Theorem 1. From our earlier analysis of the Generalized Beta 1 distribution,
we know that $\underset{p\rightarrow1^{-}}{\lim}\tfrac{1-p^{c}}{1-p}=$\linebreak{}
$\underset{p\rightarrow1^{-}}{\lim}cp^{c-1}=c>0$ by L'Hôpital's rule.
It then follows that
\[
\underset{p\rightarrow1^{-}}{\lim}f_{p|b,c>0}\left(p\right)\left(1-p\right)^{-\delta+1}=\underset{p\rightarrow1^{-}}{\lim}\dfrac{c}{\Sigma_{\textrm{B}}\left(\dfrac{1}{c},\dfrac{1}{c},b\right)}\dfrac{\left(1-p^{c}\right)^{b}}{\left(1-p\right)}\left(1-p\right)^{-\delta+1}
\]
\[
=\dfrac{c}{\Sigma_{\textrm{B}}\left(\dfrac{1}{c},\dfrac{1}{c},b\right)}\underset{p\rightarrow1^{-}}{\lim}\left(\dfrac{1-p^{c}}{1-p}\right)^{b}\left(1-p\right)^{b-\delta}
\]
\[
=\dfrac{c^{b+1}}{\Sigma_{\textrm{B}}\left(\dfrac{1}{c},\dfrac{1}{c},b\right)}\underset{p\rightarrow1^{-}}{\lim}\left(1-p\right)^{b-\delta},
\]
which is greater than 0 for all $\delta\geq b$.

We turn to Poisson mixtures in the next result, which follows directly
from Theorem 4. As in Theorem 5 and Corollary 1, part (A) considers
the case of $c\in\left(0,\infty\right)$, and part (B) the limiting
case of $c\rightarrow0^{+}$.\medskip{}

\noindent \textbf{Corollary 2:}

\noindent (A) If $X|b,c>0\sim\textrm{ZY}\left(b,c>0\right)$, then
the function
\begin{equation}
f_{\lambda|b,c>0}\left(\lambda\right)=\dfrac{c}{\Sigma_{\textrm{B}}\left(\dfrac{1}{c},\dfrac{1}{c},b\right)}{\displaystyle \int_{0}^{\infty}}\dfrac{1}{y+1}\left[1-\left(\dfrac{1}{y+1}\right)^{c}\right]^{b}\exp\left(-\lambda y\right)dy
\end{equation}
is the unique PDF satisfying $f_{X|b,c>0}^{\left(\textrm{ZY}\right)}\left(x\right)=\int_{0}^{\infty}f_{X\mid\lambda}^{\left(\textrm{P}\right)}\left(x\right)f_{\lambda|b,c>0}\left(\lambda\right)d\lambda$.

\noindent (B) If $X|b,c\rightarrow0\sim\textrm{ZY}\left(b,c\rightarrow0\right)$,
then the function
\begin{equation}
f_{\lambda|b,c\rightarrow0}\left(\lambda\right)=\dfrac{1}{\zeta\left(b+1\right)\Gamma\left(b+1\right)}{\displaystyle \int_{0}^{\infty}}\dfrac{1}{y+1}\left(\ln\left(y+1\right)\right)^{b}\exp\left(-\lambda y\right)dy
\end{equation}
is the unique PDF satisfying $f_{X|b,c\rightarrow0}^{\left(\textrm{ZY}\right)}\left(x\right)=\int_{0}^{\infty}f_{X\mid\lambda}^{\left(\textrm{P}\right)}\left(x\right)f_{\lambda|b,c\rightarrow0}\left(\lambda\right)d\lambda$.\medskip{}

\noindent \textbf{Proof:} See Subsection A.7 of the Appendix.

\subsection{Comparison with the $\textrm{Waring}\boldsymbol{\left(a,b\right)}$
Distribution}

\noindent The PMF of the two-parameter ZY distribution, (9), offers
a range of functional shapes similar to those of the two-parameter
Waring distribution, (5), and both are reasonably versatile choices
for modeling heavy-tailed loss frequencies. In fact, the two distributions'
tail behaviors are nearly identical, with tail parameter $b$ playing
essentially the same role in both cases. However, for values of $x$
close to (and including) 0, the ZY and Waring PMFs are more distinct,
with the former tending to assume flatter shapes than the latter for
any fixed value of $b$ (e.g., $\tfrac{f_{X|b,c}^{\left(\textrm{ZY}\right)}\left(0\right)}{f_{X|b,c}^{\left(\textrm{ZY}\right)}\left(1\right)}\in\left(1,2^{b+1}\right)$,
whereas $\tfrac{f_{X|a,b}^{\left(\textrm{W}\right)}\left(0\right)}{f_{X|a,b}^{\left(\textrm{W}\right)}\left(1\right)}\in\left(1,\infty\right)$).
As a consequence, the Waring distribution does not include the Zeta
model as a special case (although it does include the Yule model when
$a=1$, as noted in Subsection 3.1). Another notable difference is
that the the ZY PMF is more complex than the Waring PMF (because of
the former's normalization constant), and therefore likely to require
more complex statistical-estimation procedures. On the other hand,
the ZY distribution allows one to differentiate between the Zeta and
Yule models through hypothesis tests based on the estimated value
of the $c$ parameter.

The next result, addressing Negative Binomial mixtures, follows from
Theorem 2.\medskip{}

\noindent \textbf{Corollary 3:}

\noindent If $X|a,b\sim\textrm{Waring}\left(a,b\right)$, then:

\noindent (1) for all $r\in\left(1,\infty\right)$, the function
\[
f_{p|r>1,a,b}\left(p\right)=\dfrac{\left(r-1\right)}{\textrm{B}\left(a,b\right)\left(1-p\right)^{r}}{\displaystyle \int_{p}^{1}}\dfrac{\left(\omega-p\right)^{r-2}\left(1-\omega\right)^{b}}{\omega^{r-a}}d\omega
\]
is the unique PDF satisfying $f_{X|a,b}^{\left(\textrm{W}\right)}\left(x\right)=\int_{0}^{1}f_{X\mid r>1,p}^{\left(\textrm{NB}\right)}\left(x\right)f_{p|r>1,a,b}\left(p\right)dp$;
and

\noindent (2) for all $r\in\left(0,1\right)$ such that $r\geq a$,
the function
\[
f_{p|r<1,a,b}\left(p\right)=\dfrac{1}{\textrm{B}\left(a,b\right)\left(1-p\right)^{r}}
\]
\[
\times\left[{\displaystyle \left(r-a\right)\int_{p}^{1}}\dfrac{\left(\omega-p\right)^{r-1}\left(1-\omega\right)^{b-1}}{\omega^{r-a+1}}d\omega-\left(r-a-b\right){\displaystyle \int_{p}^{1}}\dfrac{\left(\omega-p\right)^{r-1}\left(1-\omega\right)^{b-1}}{\omega^{r-a}}d\omega\right]
\]
is the unique PDF satisfying $f_{X|a,b}^{\left(\textrm{W}\right)}\left(x\right)=\int_{0}^{1}f_{X\mid r<1,p}^{\left(\textrm{NB}\right)}\left(x\right)f_{p|r<1,a,b}\left(p\right)dp$;
and

\noindent (3) for all $r\in\left(0,1\right)$ such that $r<a$, the
function in part (2) above is a quasi-PDF satisfying $f_{X|a,b}^{\left(\textrm{W}\right)}\left(x\right)=\int_{0}^{1}f_{X\mid r<1,p}^{\left(\textrm{NB}\right)}\left(x\right)f_{p|r<1,a,b}\left(p\right)dp$,
with $f_{p|r<1,a,b}\left(p\right)\leq\ell<0$ for all $p$ in some
neighborhood of 0.\medskip{}

\noindent \textbf{Proof:} See Subsection A.8 of the Appendix.\medskip{}

The next result, addressing Poisson mixtures, follows from Theorem
4.\medskip{}

\noindent \textbf{Corollary 4:}

\noindent If $X|a,b>0\sim\textrm{Waring}\left(a,b\right)$, then the
function
\begin{equation}
f_{\lambda|a,b}\left(\lambda\right)=\dfrac{1}{\textrm{B}\left(a,b\right)}{\textstyle \int_{0}^{\infty}}\dfrac{y^{b}}{\left(y+1\right)^{a+b}}\exp\left(-y\lambda\right)dy
\end{equation}
is the unique PDF satisfying $f_{X|a,b}^{\left(\textrm{W}\right)}\left(x\right)=\int_{0}^{\infty}f_{X\mid\lambda}^{\left(\textrm{P}\right)}\left(x\right)f_{\lambda|a,b}\left(\lambda\right)d\lambda$.\medskip{}

\noindent \textbf{Proof:} See Subsection A.9 of the Appendix.

\section{Extensions and Unification}

\noindent Comparing the analytical forms of the heavy-tailed $\textrm{ZY}\left(b,c\right)$
and $\textrm{Waring}\left(a,b\right)$ PMFs is rather difficult because
the index ($x$) appears as an argument of beta functions in both
cases. Alternatively, it is quite easy to compare the associated mixing
distributions that give rise to these families as mixtures of a $\textrm{Negative Binomial}\left(r=1,p\right)$
random variable; that is, the $\textrm{Sigma-B}\left(b,c\right)$
and $\textrm{Beta}\left(a,b\right)$ PDFs, respectively. Recognizing
that
\[
f_{p|b,c}^{\left(\Sigma_{\textrm{B}}\right)}\left(p\right)=\dfrac{c}{\Sigma_{\textrm{B}}\left(\dfrac{1}{c},\dfrac{1}{c},b\right)}\dfrac{\left(1-p^{c}\right)^{b}}{\left(1-p\right)}\propto\dfrac{\left(1-p^{c}\right)^{b}}{\left(1-p\right)}
\]
and
\[
f_{p|a,b}^{\left(\textrm{B}\right)}\left(p\right)=\dfrac{1}{\textrm{B}\left(a,b\right)}p^{a-1}\left(1-p\right)^{b-1}\propto p^{a-1}\left(1-p\right)^{b-1},
\]
one can see that introducing a third parameter into each PDF can bring
their functional forms closer together in a natural way. In particular,
$f_{p|b,c}^{\left(\Sigma_{\textrm{B}}\right)}\left(p\right)$ can
be extended to what we will call the ``$\textrm{Generalized Sigma-B}\left(a,b,c\right)$''
PDF,
\begin{equation}
f_{p|a,b,c}^{\left(\textrm{G}\Sigma_{\textrm{B}}\right)}\left(p\right)=\dfrac{c}{\Sigma_{\textrm{B}}\left(\dfrac{1}{c},a,b\right)}\dfrac{p^{ca-1}\left(1-p^{c}\right)^{b}}{\left(1-p\right)}\propto\dfrac{p^{ca-1}\left(1-p^{c}\right)^{b}}{\left(1-p\right)},
\end{equation}
by inserting a factor of $p^{ca-1}$; whereas $f_{p|a,b}^{\left(\textrm{B}\right)}\left(p\right)$
can be extended to the $\textrm{Generalized Beta 1}\left(a,b,c\right)$
PDF,
\begin{equation}
f_{p|a,b,c}^{\left(\textrm{GB}1\right)}\left(p\right)=\dfrac{c}{\textrm{B}\left(a,b\right)}p^{ca-1}\left(1-p^{c}\right)^{b-1}\propto p^{ca-1}\left(1-p^{c}\right)^{b-1},
\end{equation}
by transforming $p$ to $p^{c}$.

Using (17) and (18) to construct mixtures of $\textrm{Negative Binomial}\left(r=1,p\right)$
random variables then yields the following heavy-tailed generalizations
of the ZY and Waring distributions, respectively:
\begin{equation}
f_{X|a,b,c}^{\left(\textrm{GZY}\right)}\left(x\right)=\dfrac{\Sigma_{\textrm{B}}\left(\dfrac{1}{c},\dfrac{x}{c}+a,b\right)-\Sigma_{\textrm{B}}\left(\dfrac{1}{c},\dfrac{\left(x+1\right)}{c}+a,b\right)}{\Sigma_{\textrm{B}}\left(\dfrac{1}{c},a,b\right)},
\end{equation}
to be called the ``$\textrm{Generalized ZY}\left(a,b,c\right)$''
distribution; and
\begin{equation}
f_{X|a,b,c}^{\left(\textrm{GW}2\right)}\left(x\right)=\dfrac{\textrm{B}\left(\dfrac{x}{c}+a,b\right)-\textrm{B}\left(\dfrac{\left(x+1\right)}{c}+a,b\right)}{\textrm{B}\left(a,b\right)},
\end{equation}
to be called the ``$\textrm{Generalized Waring 2}\left(a,b,c\right)$''
distribution (to distinguish it from the three-parameter Generalized
Waring distribution mentioned in Subsection 3.1). Obviously, the above
two PMFs are quite similar, with each of the three infinite series
of (19) truncated to its first term in (20).

To unify (19) and (20) into a single, four-parameter PMF, we again
turn to the associated mixing PDFs (in (17) and (18), respectively),
and note that these two cases are easily assumed into what we will
call the ``$\textrm{Hyper-Generalized Sigma-B}\left(a,b,c,d\right)$''
PDF,
\begin{equation}
f_{p|a,b,c,d}^{\left(\textrm{HG}\Sigma_{\textrm{B}}\right)}\left(p\right)=\dfrac{c}{\Sigma_{\textrm{B}}\left(\dfrac{d}{c},a,b\right)}\dfrac{p^{ca-1}\left(1-p^{c}\right)^{b}}{\left(1-p^{d}\right)},
\end{equation}
by introducing the parameter $d\in\left(0,\infty\right)$ as an exponent
of $p$ in the denominator. Using (21) to form a $\textrm{Negative Binomial}\left(r=1,p\right)$
mixture then extends the Generalized ZY and Generalized Waring 2 distributions
to the heavy-tailed model
\[
f_{X|a,b,c,d}^{\left(\textrm{HGZY}\right)}\left(x\right)=\dfrac{\Sigma_{\textrm{B}}\left(\dfrac{d}{c},\dfrac{x}{c}+a,b\right)-\Sigma_{\textrm{B}}\left(\dfrac{d}{c},\dfrac{\left(x+1\right)}{c}+a,b\right)}{\Sigma_{\textrm{B}}\left(\dfrac{d}{c},a,b\right)},
\]
to be called the ``$\textrm{Hyper-Generalized ZY}\left(a,b,c,d\right)$''
distribution.\medskip{}
\noindent \begin{center}
\includegraphics[scale=0.4]{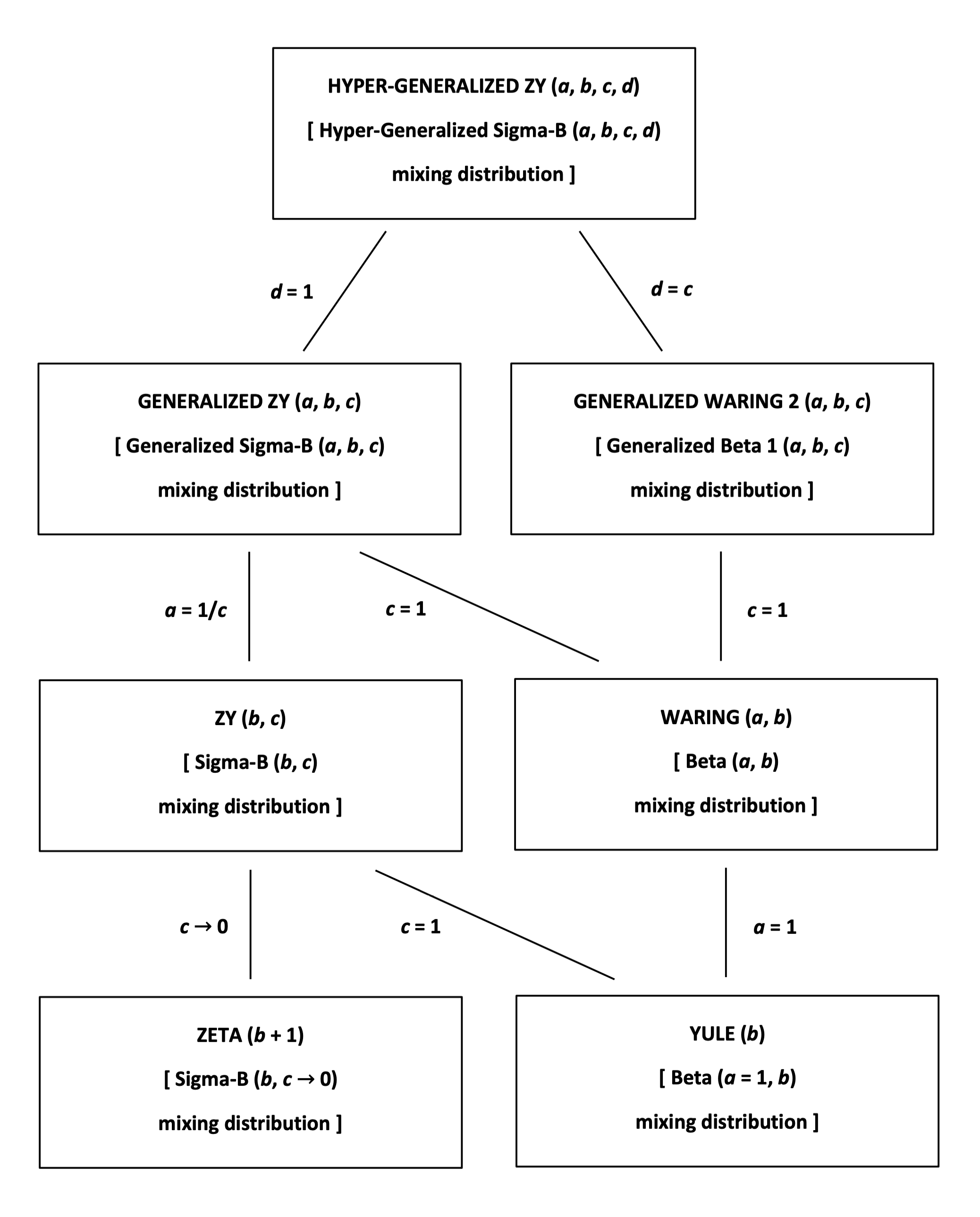}
\par\end{center}

\begin{singlespace}
\noindent \begin{center}
Figure 1. Hierarchy of Distributions within the Hyper-Generalized
ZY Family\medskip{}
\par\end{center}
\end{singlespace}

Figure 1 summarizes the parametric hierarchy among members of the
heavy-tailed Hyper-Generalized ZY family, which we will denote by
$\mathcal{HG}$. Although these relationships may be of theoretical
interest to some researchers, our purpose in formulating the $\mathcal{HG}$
family of discrete distributions \textendash{} together with the corresponding
family of continuous mixing distributions \textendash{} is more pragmatic.
Basically, we wish to provide a new and useful approach to analyzing
empirical loss frequencies in concert with conventional GLM methods.
Specifically, we would note that, after modeling frequency data with
a flexible (and possibly multi-parameter) probability distribution,
one can decompose the fitted distribution into either a Negative Binomial
or Poisson mixture via the formulas of Theorems 2 and 4, respectively.\footnote{In the Negative Binomial case, this can be done for various fixed
values of the $r$ parameter.} This approach is conceptually consistent with conventional GLM regression,
in which loss frequencies are modeled as aggregates of either Negative
Binomial or Poisson components associated with individual risk classifications,
and the exposure-weighted means of the various risk classifications
form an implicit mixing distribution. Comparing the GLM mixing distribution
with the theoretical mixing distribution of Theorem 2 or 4 thus affords
a direct means of testing the robustness of GLM methods.

\section{Application to Historical Loss Data}

\noindent In this section, the modeling approach described in Section
5 is applied to a classic set of Swedish commercial motor-vehicle
insurance loss data first studied by Hallin and Ingenbleek (1983).
Since the purpose of our analysis is primarily illustrative \textendash{}
that is, to demonstrate the usefulness of multiple-parameter loss-frequency
models based on mixtures of Poisson and Negative Binomial counts \textendash{}
we do not provide a comprehensive investigation of the data. Consequently,
certain statistical issues, such as potential collinearity among the
explanatory variables, correlation between loss frequency and loss
severity, and the presence of outliers, will not be considered.

The relevant data set is introduced in Subsection 6.1. We then carry
out the following three procedures (in Subsections 6.2-6.4, respectively):

(1) fitting the various one- to four-parameter members of $\mathcal{HG}$
to the aggregate loss-frequency counts by maximum likelihood;

(2) fitting the aggregate loss-frequency data by Poisson and Negative
Binomial GLM regression to estimate the mixing distributions implied
by the exposure-weighted means of the various risk classifications;
and

(3) fitting the distributions of parameter estimates generated by
the Poisson and Negative Binomial GLM regressions by the theoretical
mixing distributions associated with the members of $\mathcal{HG}$.

\subsection{Swedish Commercial Motor-Vehicle Data}

\noindent Sweden's 1977 portfolio of commercial motor-vehicle third-party
insurance loss data was introduced to the actuarial literature by
Hallin and Ingenbleek (1983) for purposes of investigating a new methodology
for risk-based pricing. These data then were included in a compilation
of data sets by Andrews and Herzberg (1985), and subsequently used
by numerous researchers (see, e.g.: Smyth and Jørgensen, 2002; Frees,
2010; Pigeon and Denuit, 2011; Pan, Soo, and Pooi, 2014; Dunn and
Smyth, 2018; and Zhang, 2021). One important characteristic of the
Swedish commercial motor-vehicle portfolio is the heavy tail of its
empirical loss-frequency distribution, which appears not to have been
discussed in prior articles.

The Swedish data are broken down by policyholder and motor-vehicle
characteristics, with 2182 subportfolios (line items) in total. Each
line item, $i=1,2,\ldots,2182$, includes seven fields:

(a) number of kilometers driven (as one of 5 ranges: $\left[0,\:1000\right)$,
$\left[1000,\:15,000\right)$, $\left[15,000,\:20,000\right)$, $\left[20,000,\:25,000\right)$,
$\left[25,0000,\:\infty\right)$);

(b) geographical zone (as one of 7 regions\footnote{See Hallin and Ingenbleek (1983) for descriptions of the geographical
zones and motor-vehicle makes.});

(c) no-claim bonus status ($=\min\left\{ 1+\textrm{number of years since last claim},\:7\right\} );$

(d) motor-vehicle make (as one of 9 categories: 8 specific models
plus 1 ``all other'');

(e) number of exposures (insured-vehicle policy years), $E_{i}$;

(f) number of loss events (recorded as claims), $X_{i}$; and

(g) total losses (in Swedish Krona);

\noindent and is formed by a unique combination of risk classifications
(items (a)-(d)) aggregated over all insurance companies.

Summary statistics from the empirical distribution of the $X_{i}$
are presented in Table 1, and include notably large coefficients of
skewness and kurtosis. The high degree of positive skewness is clear
from the plot of the empirical histogram in Figure 2, and the heavy
right tail confirmed by the plot of the empirical hazard function
in Figure 3, which tends to 0 for large values in the sample space.\medskip{}
\begin{singlespace}
\noindent \begin{center}
Table 1. Summary Statistics of Swedish Motor-Vehicle Frequencies,
$X_{i}$
\par\end{center}

\noindent \begin{center}
\begin{tabular}{|c|c|c|c|c|c|c|}
\hline 
\textbf{\small{}No. of Obs.} & \textbf{\small{}Minimum} & \textbf{\small{}Maximum} & \textbf{\small{}\quad{}Mean\quad{}} & \textbf{\small{}\enskip{}Std. Dev.\enskip{}} & \textbf{\small{}Skewness} & \textbf{\small{}Kurtosis}\tabularnewline
\hline 
{\small{}2182} & {\small{}0} & {\small{}3338} & {\small{}51.87} & {\small{}201.71} & {\small{}8.57} & {\small{}96.36}\tabularnewline
\hline 
\end{tabular}\medskip{}
\par\end{center}

\noindent \begin{center}
\includegraphics[scale=0.7]{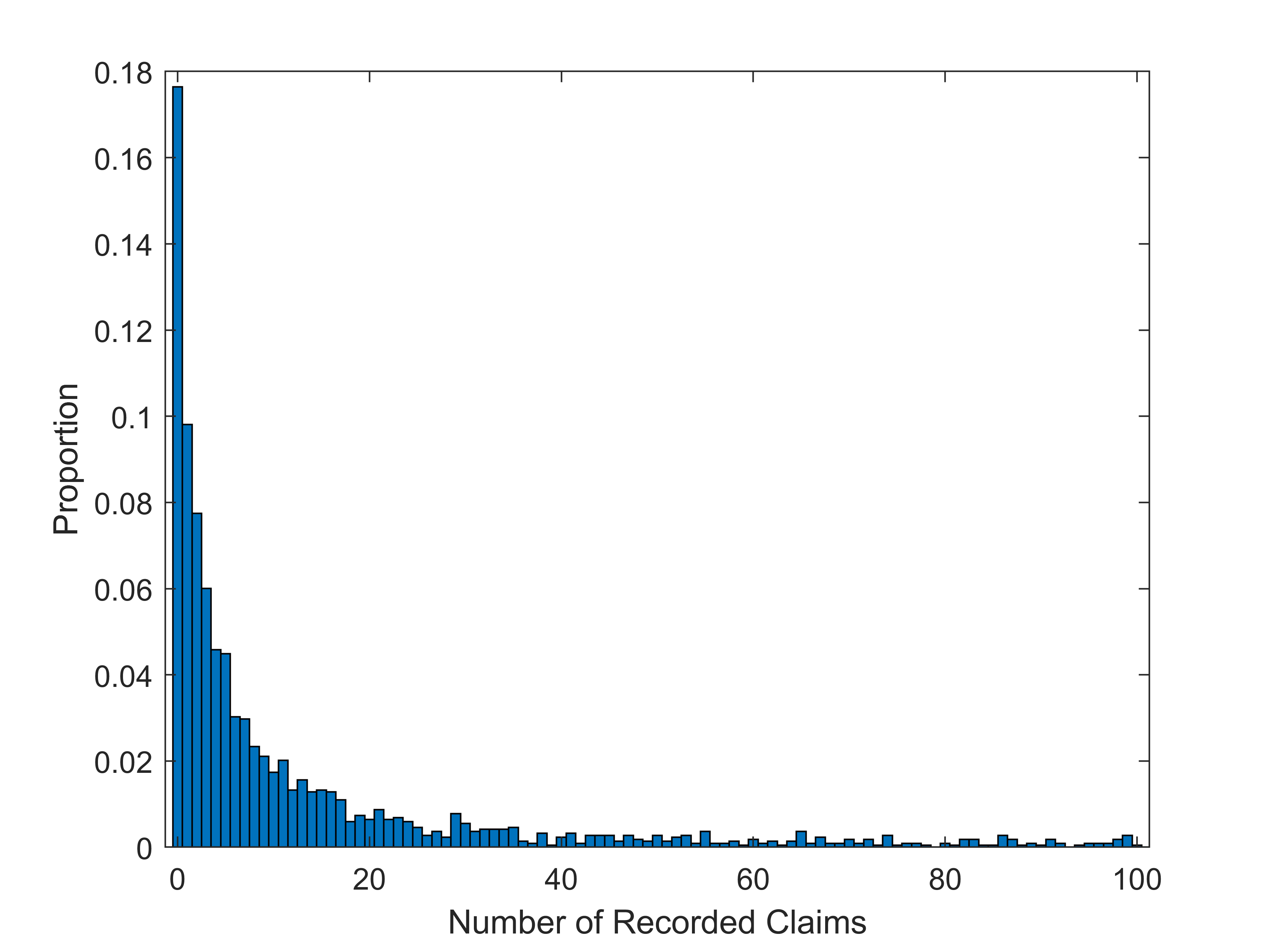}
\par\end{center}

\noindent \begin{center}
Figure 2. Swedish Motor-Vehicle Frequencies, Empirical Histogram\medskip{}
\includegraphics[scale=0.7]{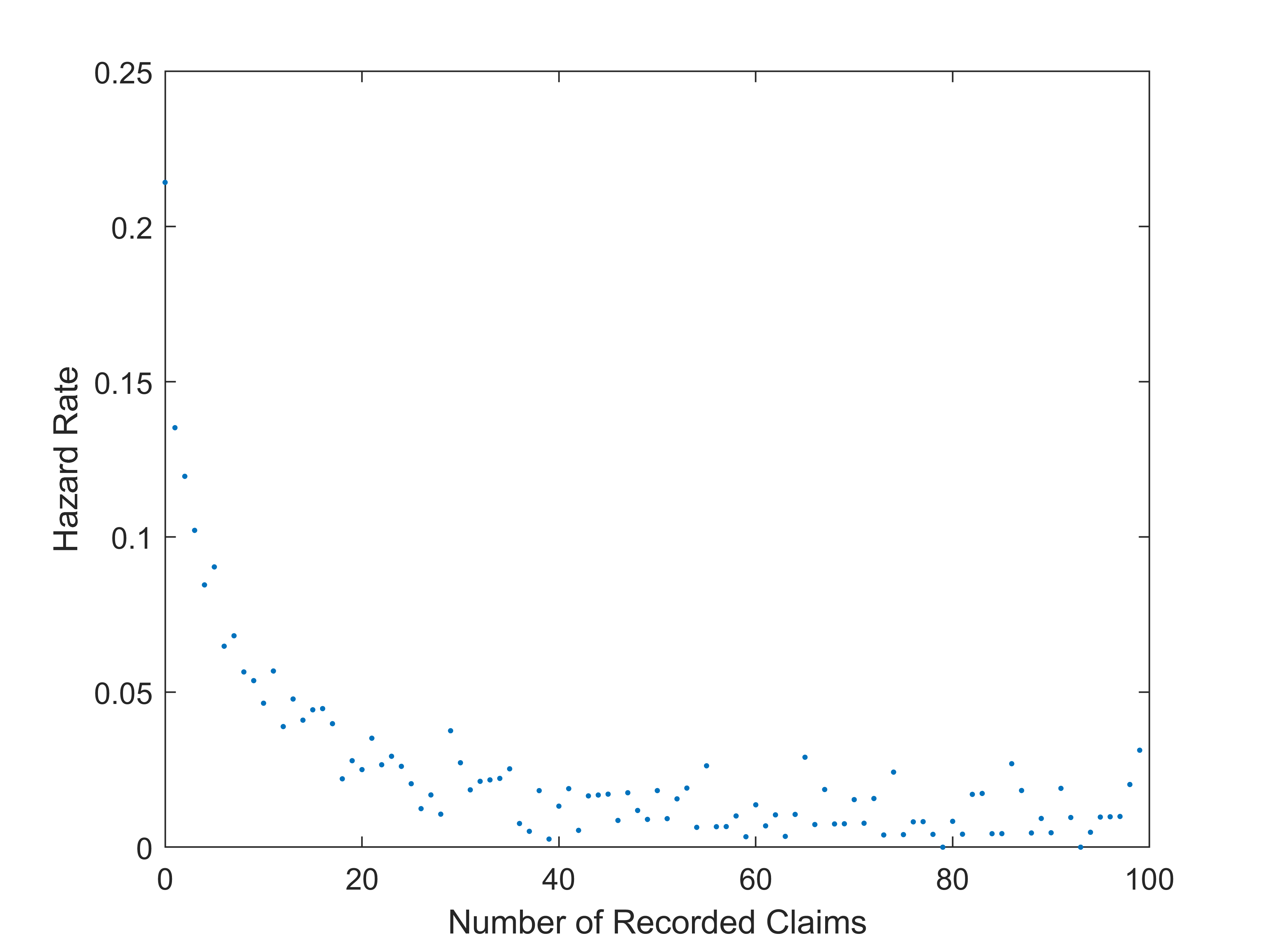}
\par\end{center}

\noindent \begin{center}
Figure 3. Swedish Motor-Vehicle Frequencies, Empirical Hazard Function\medskip{}
\par\end{center}
\end{singlespace}

The observed tail characteristics suggest that the frequency data
may be well modeled by one or more members of $\mathcal{HG}$. However,
it is important to note that much of the tail behavior is explained
by heavy tails in the empirical distribution of exposures. This is
shown in Table 2, Figures 4, and Figure 5, which demonstrate skewness
and heavy-tail properties comparable to those of the empirical frequency
distribution. In the context of the present analysis, this means that:
(1) when fitting a member of $\mathcal{HG}$ to the data, it is likely
that the underlying exposure characteristics strongly influence the
relevant mixing distribution; and (2) in applying GLM regression,
the exposures constitute a natural offset.\medskip{}
\begin{singlespace}
\noindent \begin{center}
Table 2. Summary Statistics of Swedish Motor-Vehicle Exposures, $E_{i}$
\par\end{center}

\noindent \begin{center}
\begin{tabular}{|c|c|c|c|c|c|c|}
\hline 
\textbf{\small{}No. of Obs.} & \textbf{\small{}Minimum} & \textbf{\small{}Maximum} & \textbf{\small{}\quad{}Mean\quad{}} & \textbf{\small{}\enskip{}Std. Dev.\enskip{}} & \textbf{\small{}Skewness} & \textbf{\small{}Kurtosis}\tabularnewline
\hline 
{\small{}2182} & {\small{}0.01} & {\small{}127,687.27} & {\small{}1092.20} & {\small{}5661.16} & {\small{}13.96} & {\small{}253.49}\tabularnewline
\hline 
\end{tabular}\medskip{}
\includegraphics[scale=0.7]{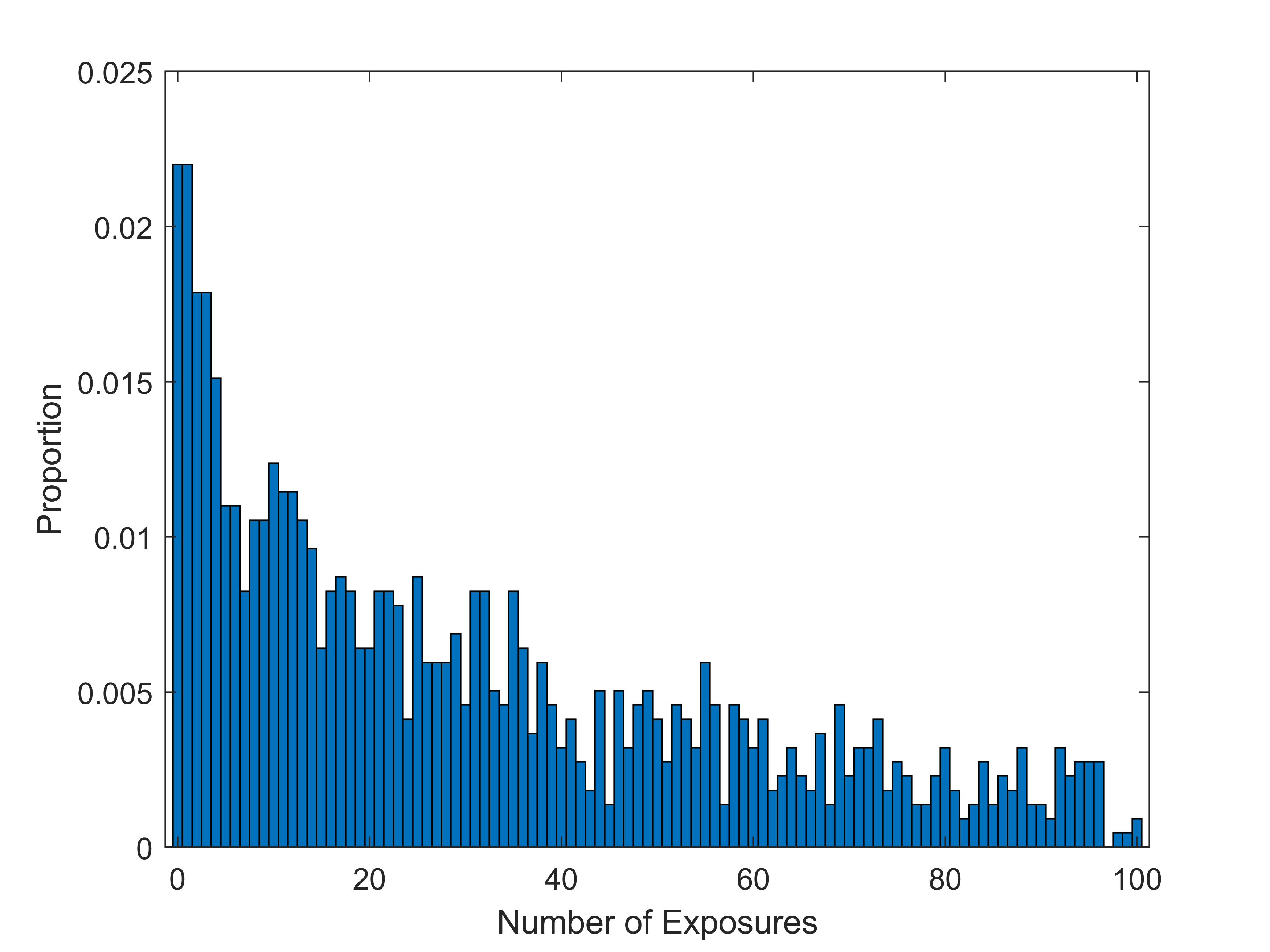}
\par\end{center}

\noindent \begin{center}
Figure 4. Swedish Motor-Vehicle Exposures, Empirical Histogram\medskip{}
\includegraphics[scale=0.7]{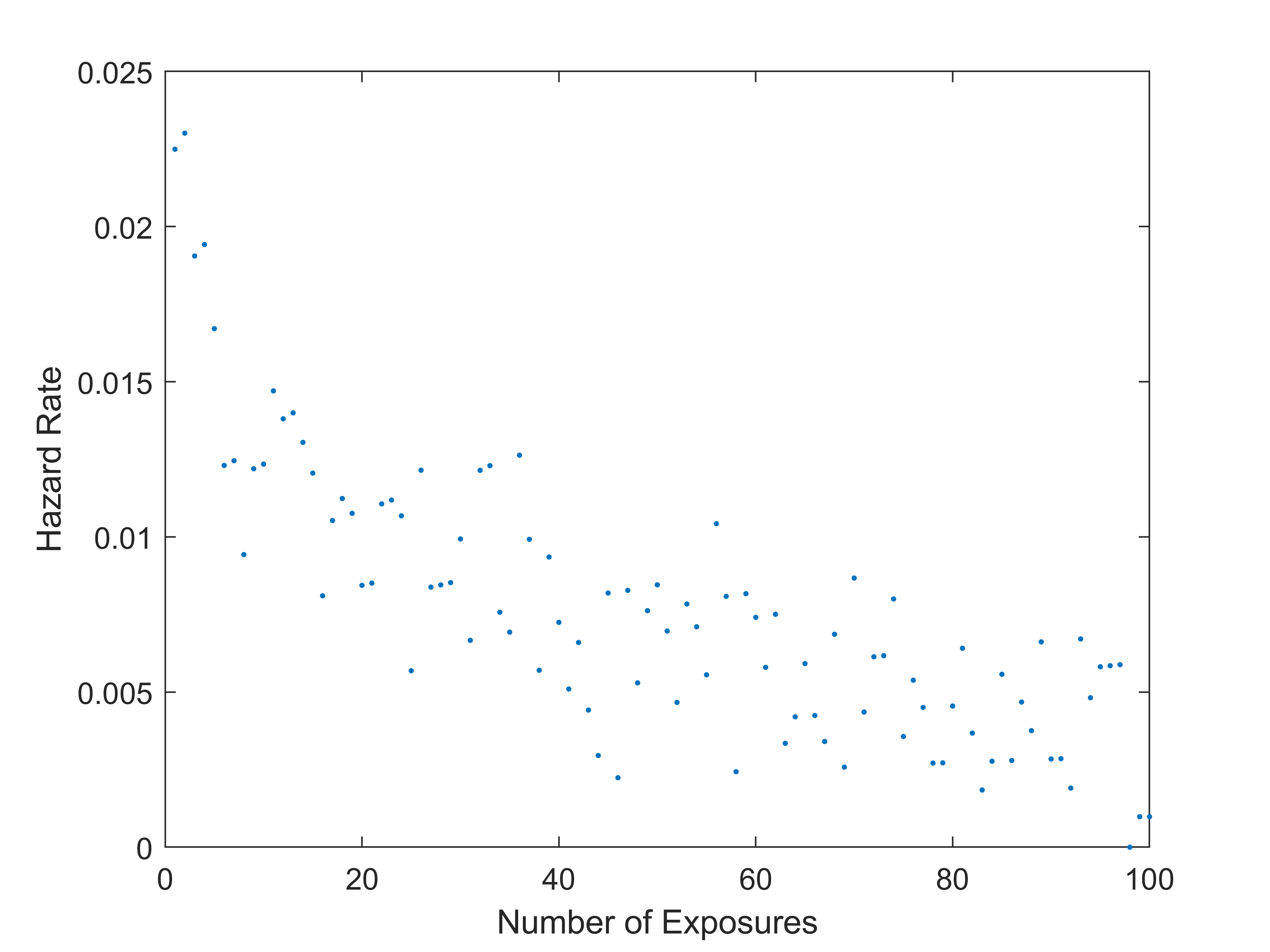}
\par\end{center}

\noindent \begin{center}
Figure 5. Swedish Motor-Vehicle Exposures, Empirical Hazard Function
\par\end{center}
\end{singlespace}

\subsection{Fitting $\mathcal{HG}$ Frequency Models}

\noindent We fit the seven members of $\mathcal{HG}$ to the motor-vehicle
frequencies, $X_{i}$, using the method of maximum likelihood. As
shown in Table 3, this provides useful results in all but one case
\textendash{} that of the three-parameter $\textrm{GW2}\left(a,b,c\right)$
distribution \textendash{} for which the likelihood function possesses
no interior maximum.\medskip{}
\begin{singlespace}
\noindent \begin{center}
Table 3. Maximum-Likelihood Estimation of $\mathcal{HG}$ Parameters
\par\end{center}

\noindent \begin{center}
\begin{tabular}{|c|c|c|c|c|c|}
\hline 
\textbf{\small{}Distribution} & \textbf{\small{}$\boldsymbol{\hat{a}}$ } & \textbf{\small{}$\boldsymbol{\hat{b}}$ } & \textbf{\small{}$\boldsymbol{\hat{c}}$ } & \textbf{\small{}$\boldsymbol{\hat{d}}$ } & \textbf{\small{}Log(Likelihood)}\tabularnewline
\hline 
{\small{}$\textrm{HGZY}\left(a,b,c,d\right)$} & {\small{}0.0049} & {\small{}3.3112} & {\small{}939.1870} & {\small{}70.0691} & {\small{}-8668.78}\tabularnewline
 & {\small{}(0.0040){*}} & {\small{}(2.8519)} & {\small{}(809.2498)} & {\small{}(23.8098)} & \tabularnewline
\hline 
{\small{}$\textrm{GZY}\left(a,b,c\right)$} & {\small{}0.0727} & {\small{}0.8997} & {\small{}23.6117} &  & {\small{}-8675.28}\tabularnewline
 & {\small{}(0.0290)} & {\small{}(0.0670)} & {\small{}(6.9803)} &  & \tabularnewline
\hline 
{\small{}$\textrm{GW2}\left(a,b,c\right)$} & {\small{}$---$} & {\small{}$---$} & {\small{}$---$} &  & \emph{\small{}Fails to converge}\tabularnewline
 &  &  &  &  & \tabularnewline
\hline 
{\small{}$\textrm{ZY}\left(b,c\right)$} &  & {\small{}1.0909} & {\small{}60.8621} &  & {\small{}-8690.72}\tabularnewline
 &  & {\small{}(0.0919)} & {\small{}(10.1789)} &  & \tabularnewline
\hline 
{\small{}$\textrm{Waring}\left(a,b\right)$} & {\small{}3.9178} & {\small{}0.7431} &  &  & {\small{}-8682.03}\tabularnewline
 & {\small{}(0.2755)} & {\small{}(0.0287)} &  &  & \tabularnewline
\hline 
{\small{}$\textrm{Zeta}\left(b+1\right)$} &  & {\small{}0.3733} &  &  & {\small{}-8927.62}\tabularnewline
 &  & {\small{}(0.0086)} &  &  & \tabularnewline
\hline 
{\small{}$\textrm{Yule}\left(b\right)$} &  & {\small{}0.4138} &  &  & {\small{}-8876.67}\tabularnewline
 &  & {\small{}(0.0100)} &  &  & \tabularnewline
\hline 
\end{tabular}
\par\end{center}
\end{singlespace}

\begin{singlespace}
\noindent \begin{flushleft}
{\small{}{*} Estimated standard errors of parameter estimators are
shown in parentheses.}\medskip{}
\par\end{flushleft}
\end{singlespace}

To assess which (if any) of the $\mathcal{HG}$ distributions fit
the motor-vehicle data well, we first employ the maximum-likelihood
parameter estimates to compute minimum-$\chi^{2}$ statistics for
each of the feasible six members. Table 4 shows that, at commonly
used levels of significance, the minimum-$\chi^{2}$ test rejects
the null hypothesis (that the motor-vehicle frequency data come from
the indicated distribution) for the $\textrm{ZY}\left(b,c\right)$,
$\textrm{Zeta}\left(b+1\right)$, and $\textrm{Yule}\left(b\right)$
models. To compare the relative suitability of the remaining three
distributions, we next consider the Akaike information criterion (AIC)
and Bayesian information criterion (BIC), provided in the two rightmost
columns of the same table. Both of these measures suggest that, even
after including explicit penalties for additional parameters, the
four-parameter $\textrm{HGZY}\left(a,b,c,d\right)$ is preferred to
the three-parameter $\textrm{GZY}\left(a,b,c\right)$, which in turn
is preferred to the two-parameter $\textrm{Waring}\left(a,b\right)$.\pagebreak{}
\begin{singlespace}
\noindent \begin{center}
Table 4. Goodness-of-Fit Measures for $\mathcal{HG}$ Distributions
\par\end{center}

\noindent \begin{center}
\begin{tabular}{|c|c|c|c||c|c|}
\hline 
\textbf{\small{}Distribution} & \textbf{\small{}Min.-}{\small{}$\boldsymbol{\chi^{2}}$}\textbf{\small{}
Statistic} & \textbf{\small{}Degrees of Freedom} & \textbf{\small{}$\boldsymbol{p}$-Value} & \textbf{\small{}AIC} & \textbf{\small{}BIC}\tabularnewline
\hline 
{\small{}$\textrm{HGZY}\left(a,b,c,d\right)$} & {\small{}24.57} & {\small{}29{*} - 4 - 1 = 24} & {\small{}0.4291} & {\small{}17,345.56} & {\small{}17,368.32}\tabularnewline
\hline 
{\small{}$\textrm{GZY}\left(a,b,c\right)$} & {\small{}26.88} & {\small{}32 - 3 - 1 = 28} & {\small{}0.5246} & {\small{}17,356.56} & {\small{}17,373.62}\tabularnewline
\hline 
{\small{}$\textrm{ZY}\left(b,c\right)$} & {\small{}64.65} & {\small{}31 - 2 - 1 = 28} & {\small{}0.0001} &  & \tabularnewline
\hline 
{\small{}$\textrm{Waring}\left(a,b\right)$} & {\small{}27.71} & {\small{}30 - 2 - 1 = 27} & {\small{}0.4260} & {\small{}17,368.06} & {\small{}17,379.44}\tabularnewline
\hline 
{\small{}$\textrm{Zeta}\left(b+1\right)$} & {\small{}327.88} & {\small{}22 - 1 - 1 = 20} & {\small{}\textless{} 0.0001} &  & \tabularnewline
\hline 
{\small{}$\textrm{Yule}\left(b\right)$} & {\small{}273.67} & {\small{}23 - 1 - 1 = 21} & {\small{}\textless{} 0.0001} &  & \tabularnewline
\hline 
\end{tabular}
\par\end{center}
\end{singlespace}

\begin{singlespace}
\noindent \begin{flushleft}
{\small{}{*} In computing the minimum-$\chi^{2}$ statistic for a
given distribution, the cells consist of all nonnegative integers
$x$ such that $2182\times\Pr\left\{ X=x\right\} \geq10$, plus 1
additional cell for all remaining integers $x$.}
\par\end{flushleft}{\small \par}
\end{singlespace}

\subsection{Applying GLM Regression}

\noindent Distributional modeling of portfolio loss frequencies, such
as that performed in the previous subsection, offers no direct information
about the loss propensities of individual risk classifications. For
this reason, actuaries typically work with GLM or similar regression
methods that permit the estimation of each risk classification's contribution
to overall losses. Although a wide array of sophisticated extensions
and generalizations of GLM regression are now available (see, e.g.,
Goldburd, Khare, Tevet, and Guller, 2020), the basic GLM framework
for expected loss-frequency with Poisson and/or Negative Binomial
observations remains an integral part of actuarial practice. Therefore,
we illustrate the usefulness of our mixing-distribution approach by
fitting the Swedish motor-vehicle data with a simple GLM with natural-log
link function and exposures as offset,
\begin{equation}
\ln\left(\mu_{i}\right)=\ln\left(E_{i}\right)+\beta_{0}+\mathbf{Y}_{i}^{\left(\textrm{KILO}\right)}\boldsymbol{\beta}^{\left(\textrm{KILO}\right)}+\mathbf{Y}_{i}^{\left(\textrm{ZONE}\right)}\boldsymbol{\beta}^{\left(\textrm{ZONE}\right)}+\mathbf{Y}{}_{i}^{\left(\textrm{BONUS}\right)}\boldsymbol{\beta}^{\left(\textrm{BONUS}\right)}+\mathbf{Y}_{i}^{\left(\textrm{MAKE}\right)}\boldsymbol{\beta}^{\left(\textrm{MAKE}\right)},
\end{equation}
where:

$\mathbf{Y}_{i}^{\left(\textrm{KILO}\right)}$, $\mathbf{Y}_{i}^{\left(\textrm{ZONE}\right)}$,
$\mathbf{Y}_{i}^{\left(\textrm{BONUS}\right)}$, and $\mathbf{Y}_{i}^{\left(\textrm{MAKE}\right)}$
denote line item $i$'s four fields (from (a)-(d) from Subsection
6.1) encoded as matrices of indicator variables;

$\mu_{i}=E\left[X_{i}\mid E_{i},\mathbf{Y}_{i}^{\left(\textrm{KILO}\right)},\mathbf{Y}_{i}^{\left(\textrm{ZONE}\right)},\mathbf{Y}_{i}^{\left(\textrm{BONUS}\right)},\mathbf{Y}_{i}^{\left(\textrm{MAKE}\right)}\right]$;
and either

(I) $X_{i}\mid E_{i},\mathbf{Y}_{i}^{\left(\textrm{KILO}\right)},\mathbf{Y}_{i}^{\left(\textrm{ZONE}\right)},\mathbf{Y}_{i}^{\left(\textrm{BONUS}\right)},\mathbf{Y}_{i}^{\left(\textrm{MAKE}\right)}\sim\textrm{independent Poisson}\left(\lambda_{i}\right)$
or

(II) $X_{i}\mid E_{i},\mathbf{Y}_{i}^{\left(\textrm{KILO}\right)},\mathbf{Y}_{i}^{\left(\textrm{ZONE}\right)},\mathbf{Y}_{i}^{\left(\textrm{BONUS}\right)},\mathbf{Y}_{i}^{\left(\textrm{MAKE}\right)}\sim\textrm{independent Negative Binomial}\left(r,p_{i}\right)$.

\noindent For simplicity, the formulation with Poisson observations
will be called ``Model I'', and that with Negative Binomial observations
``Model II''.

From (22), it is clear that both Models I and II include the full
set of four risk-classification variables described in Subsection
6.1. For completeness, we ran GLM regressions using all subsets of
these four variables, and found that both the AIC and BIC were minimized
under the full model for both the Poisson and Negative Binomial cases.
Therefore, only the full-model results are presented in Tables 5 and
6.\medskip{}
\begin{singlespace}
\noindent \begin{center}
Table 5. Model I (Poisson) Regression Results
\par\end{center}

\noindent \begin{center}
\begin{tabular}{|c|c|c|r@{\extracolsep{0pt}.}l|}
\hline 
\textbf{\small{}Log(Likelihood)} & \textbf{\small{}Residual-Deviance} & \textbf{\small{}Degrees of Freedom} & \multicolumn{2}{c|}{\textbf{\small{}$\boldsymbol{p}$-Value}}\tabularnewline
 & {\small{}$\boldsymbol{\chi^{2}}$}\textbf{\small{} Statistic} &  & \multicolumn{2}{c|}{}\tabularnewline
\hline 
{\small{}-5302.00} & {\small{}2966.12} & {\small{}2182 - 25 = 2157} & {\small{}\textless{} 0}&{\small{}0001}\tabularnewline
\hline 
\hline 
\textbf{\small{}Parameter} & \textbf{\small{}Estimate} & \textbf{\small{}Standard Error} & \multicolumn{2}{c|}{\textbf{\small{}$\boldsymbol{z}$-Value}}\tabularnewline
\hline 
{\small{}$\beta_{0}$} & {\small{}-1.8128} & {\small{}0.0138} & {\small{}-131}&{\small{}7754}\tabularnewline
\hline 
{\small{}$\beta_{2}^{\left(\textrm{KILO}\right)}$} & {\small{}0.2126} & {\small{}0.0075} & {\small{}28}&{\small{}2549}\tabularnewline
\hline 
{\small{}$\beta_{3}^{\left(\textrm{KILO}\right)}$} & {\small{}0.3202} & {\small{}0.0087} & {\small{}36}&{\small{}9737}\tabularnewline
\hline 
{\small{}$\beta_{4}^{\left(\textrm{KILO}\right)}$} & {\small{}0.4047} & {\small{}0.0121} & {\small{}33}&{\small{}5715}\tabularnewline
\hline 
{\small{}$\beta_{5}^{\left(\textrm{KILO}\right)}$} & {\small{}0.5760} & {\small{}0.0128} & {\small{}44}&{\small{}8916}\tabularnewline
\hline 
{\small{}$\beta_{2}^{\left(\textrm{ZONE}\right)}$} & {\small{}-0.2382} & {\small{}0.0095} & {\small{}-25}&{\small{}0820}\tabularnewline
\hline 
{\small{}$\beta_{3}^{\left(\textrm{ZONE}\right)}$} & {\small{}-0.3864} & {\small{}0.0097} & {\small{}-39}&{\small{}9587}\tabularnewline
\hline 
{\small{}$\beta_{4}^{\left(\textrm{ZONE}\right)}$} & {\small{}-0.5819} & {\small{}0.0087} & {\small{}-67}&{\small{}2428}\tabularnewline
\hline 
{\small{}$\beta_{5}^{\left(\textrm{ZONE}\right)}$} & {\small{}-0.3261} & {\small{}0.0145} & {\small{}-22}&{\small{}4455}\tabularnewline
\hline 
{\small{}$\beta_{6}^{\left(\textrm{ZONE}\right)}$} & {\small{}-0.5262} & {\small{}0.0119} & {\small{}-44}&{\small{}3086}\tabularnewline
\hline 
{\small{}$\beta_{7}^{\left(\textrm{ZONE}\right)}$} & {\small{}-0.7310} & {\small{}0.0407} & {\small{}-17}&{\small{}9617}\tabularnewline
\hline 
{\small{}$\beta_{2}^{\left(\textrm{BONUS}\right)}$} & {\small{}-0.4790} & {\small{}0.0121} & {\small{}-39}&{\small{}6069}\tabularnewline
\hline 
{\small{}$\beta_{3}^{\left(\textrm{BONUS}\right)}$} & {\small{}-0.6932} & {\small{}0.0135} & {\small{}-51}&{\small{}3160}\tabularnewline
\hline 
{\small{}$\beta_{4}^{\left(\textrm{BONUS}\right)}$} & {\small{}-0.8274} & {\small{}0.0146} & {\small{}-56}&{\small{}7348}\tabularnewline
\hline 
{\small{}$\beta_{5}^{\left(\textrm{BONUS}\right)}$} & {\small{}-0.9256} & {\small{}0.0140} & {\small{}-66}&{\small{}2689}\tabularnewline
\hline 
{\small{}$\beta_{6}^{\left(\textrm{BONUS}\right)}$} & {\small{}-0.9935} & {\small{}0.0116} & {\small{}-85}&{\small{}4294}\tabularnewline
\hline 
{\small{}$\beta_{7}^{\left(\textrm{BONUS}\right)}$} & {\small{}-1.3274} & {\small{}0.0087} & {\small{}-152}&{\small{}8445}\tabularnewline
\hline 
{\small{}$\beta_{2}^{\left(\textrm{MAKE}\right)}$} & {\small{}0.0762 } & {\small{}0.0212} & {\small{}3}&{\small{}5898}\tabularnewline
\hline 
{\small{}$\beta_{3}^{\left(\textrm{MAKE}\right)}$} & {\small{}-0.2474 } & {\small{}0.0251} & {\small{}-9}&{\small{}8594}\tabularnewline
\hline 
{\small{}$\beta_{4}^{\left(\textrm{MAKE}\right)}$} & {\small{}-0.6535} & {\small{}0.0242} & {\small{}-27}&{\small{}0218}\tabularnewline
\hline 
{\small{}$\beta_{5}^{\left(\textrm{MAKE}\right)}$} & {\small{}0.1549} & {\small{}0.0202} & {\small{}7}&{\small{}6564}\tabularnewline
\hline 
{\small{}$\beta_{6}^{\left(\textrm{MAKE}\right)}$} & {\small{}-0.3356} & {\small{}0.0174} & {\small{}-19}&{\small{}3139}\tabularnewline
\hline 
{\small{}$\beta_{7}^{\left(\textrm{MAKE}\right)}$} & {\small{}-0.0559} & {\small{}0.0233} & {\small{}-2}&{\small{}3965}\tabularnewline
\hline 
{\small{}$\beta_{8}^{\left(\textrm{MAKE}\right)}$} & {\small{}-0.0439} & {\small{}0.0316} & {\small{}-1}&{\small{}3901}\tabularnewline
\hline 
{\small{}$\beta_{9}^{\left(\textrm{MAKE}\right)}$} & {\small{}-0.0681} & {\small{}0.0100} & {\small{}-6}&{\small{}8356}\tabularnewline
\hline 
\end{tabular}\pagebreak{}
\par\end{center}

\noindent \begin{center}
Table 6. Model II (Negative Binomial) Regression Results
\par\end{center}

\noindent \begin{center}
\begin{tabular}{|c|c|c|r@{\extracolsep{0pt}.}l|}
\hline 
\textbf{\small{}Log(Likelihood)} & \textbf{\small{}Residual-Deviance} & \textbf{\small{}Degrees of Freedom} & \multicolumn{2}{c|}{\textbf{\small{}$\boldsymbol{p}$-Value}}\tabularnewline
 & {\small{}$\boldsymbol{\chi^{2}}$}\textbf{\small{} Statistic} &  & \multicolumn{2}{c|}{}\tabularnewline
\hline 
{\small{}-5163.80} & {\small{}2229.91} & {\small{}2182 - 26 = 2156} & {\small{}0}&{\small{}1307}\tabularnewline
\hline 
\hline 
\textbf{\small{}Parameter} & \textbf{\small{}Estimate} & \textbf{\small{}Standard Error} & \multicolumn{2}{c|}{\textbf{\small{}$\boldsymbol{z}$-Value}}\tabularnewline
\hline 
{\small{}$r$} & {\small{}110.5986} & {\small{}14.8104} & {\small{}7}&{\small{}4676}\tabularnewline
\hline 
{\small{}$\beta_{0}$} & {\small{}-1.7822} & {\small{}0.0238} & {\small{}-75}&{\small{}0413}\tabularnewline
\hline 
{\small{}$\beta_{2}^{\left(\textrm{KILO}\right)}$} & {\small{}0.1885} & {\small{}0.0150} & {\small{}12}&{\small{}5337}\tabularnewline
\hline 
{\small{}$\beta_{3}^{\left(\textrm{KILO}\right)}$} & {\small{}0.2753} & {\small{}0.0162} & {\small{}17}&{\small{}0058}\tabularnewline
\hline 
{\small{}$\beta_{4}^{\left(\textrm{KILO}\right)}$} & {\small{}0.3521} & {\small{}0.0193} & {\small{}18}&{\small{}2378}\tabularnewline
\hline 
{\small{}$\beta_{5}^{\left(\textrm{KILO}\right)}$} & {\small{}0.5167} & {\small{}0.0201} & {\small{}25}&{\small{}7438}\tabularnewline
\hline 
{\small{}$\beta_{2}^{\left(\textrm{ZONE}\right)}$} & {\small{}-0.2242} & {\small{}0.0179} & {\small{}-12}&{\small{}5241}\tabularnewline
\hline 
{\small{}$\beta_{3}^{\left(\textrm{ZONE}\right)}$} & {\small{}-0.3828} & {\small{}0.0181} & {\small{}-21}&{\small{}2094}\tabularnewline
\hline 
{\small{}$\beta_{4}^{\left(\textrm{ZONE}\right)}$} & {\small{}-0.5559} & {\small{}0.0170} & {\small{}-32}&{\small{}7396}\tabularnewline
\hline 
{\small{}$\beta_{5}^{\left(\textrm{ZONE}\right)}$} & {\small{}-0.3386} & {\small{}0.0225} & {\small{}-15}&{\small{}0191}\tabularnewline
\hline 
{\small{}$\beta_{6}^{\left(\textrm{ZONE}\right)}$} & {\small{}-0.5225} & {\small{}0.0200} & {\small{}-26}&{\small{}1238}\tabularnewline
\hline 
{\small{}$\beta_{7}^{\left(\textrm{ZONE}\right)}$} & {\small{}-0.7307} & {\small{}0.0462} & {\small{}-15}&{\small{}8003}\tabularnewline
\hline 
{\small{}$\beta_{2}^{\left(\textrm{BONUS}\right)}$} & {\small{}-0.4428} & {\small{}0.0214} & {\small{}-20}&{\small{}7266}\tabularnewline
\hline 
{\small{}$\beta_{3}^{\left(\textrm{BONUS}\right)}$} & {\small{}-0.6805} & {\small{}0.0224} & {\small{}-30}&{\small{}3736}\tabularnewline
\hline 
{\small{}$\beta_{4}^{\left(\textrm{BONUS}\right)}$} & {\small{}-0.8204} & {\small{}0.0231} & {\small{}-35}&{\small{}4871}\tabularnewline
\hline 
{\small{}$\beta_{5}^{\left(\textrm{BONUS}\right)}$} & {\small{}-0.9191} & {\small{}0.0224} & {\small{}-41}&{\small{}0399}\tabularnewline
\hline 
{\small{}$\beta_{6}^{\left(\textrm{BONUS}\right)}$} & {\small{}-0.9931} & {\small{}0.0203} & {\small{}-48}&{\small{}8821}\tabularnewline
\hline 
{\small{}$\beta_{7}^{\left(\textrm{BONUS}\right)}$} & {\small{}-1.3259} & {\small{}0.0177} & {\small{}-74}&{\small{}8806}\tabularnewline
\hline 
{\small{}$\beta_{2}^{\left(\textrm{MAKE}\right)}$} & {\small{}0.0673} & {\small{}0.0256} & {\small{}2}&{\small{}6263}\tabularnewline
\hline 
{\small{}$\beta_{3}^{\left(\textrm{MAKE}\right)}$} & {\small{}-0.2352} & {\small{}0.0294} & {\small{}-7}&{\small{}9859}\tabularnewline
\hline 
{\small{}$\beta_{4}^{\left(\textrm{MAKE}\right)}$} & {\small{}-0.6836} & {\small{}0.0293} & {\small{}-23}&{\small{}3053}\tabularnewline
\hline 
{\small{}$\beta_{5}^{\left(\textrm{MAKE}\right)}$} & {\small{}0.1523} & {\small{}0.0245} & {\small{}6}&{\small{}2119}\tabularnewline
\hline 
{\small{}$\beta_{6}^{\left(\textrm{MAKE}\right)}$} & {\small{}-0.3633} & {\small{}0.0221} & {\small{}-16}&{\small{}4340}\tabularnewline
\hline 
{\small{}$\beta_{7}^{\left(\textrm{MAKE}\right)}$} & {\small{}-0.0791} & {\small{}0.0277} & {\small{}-2}&{\small{}8564}\tabularnewline
\hline 
{\small{}$\beta_{8}^{\left(\textrm{MAKE}\right)}$} & {\small{}-0.0411} & {\small{}0.0352} & {\small{}-1}&{\small{}1695}\tabularnewline
\hline 
{\small{}$\beta_{9}^{\left(\textrm{MAKE}\right)}$} & {\small{}-0.0903} & {\small{}0.0157} & {\small{}-5}&{\small{}7576}\tabularnewline
\hline 
\end{tabular}\medskip{}
\medskip{}
\medskip{}
\par\end{center}
\end{singlespace}

The contents of these two tables show that Model II affords a better
fit to the motor-vehicle data than Model I. This is most apparent
from comparing the respective $p$-values, since Model I's value (\textless{}
0.0001) strongly supports rejecting (22) with the Poisson assumption,
whereas Model II's value (0.1307) does not support rejecting (22)
with the Negative Binomial assumption at common levels of significance.
Furthermore, a close inspection of the $z$-values associated with
the individual risk classifications reveals these quantities to be
substantially more variable in Model I than Model II, suggesting that
the former model gives disproportionate leverage to risk classifications
with less overdispersion.

In the next subsection, we will use these models to analyze the mixing
distributions implied by the exposure-weighted means of the various
risk classifications, providing insights into the relationship between
the mixing distributions and overall model fit. For both Models I
and II, the set of exposure-weighted risk-classification means is
given by
\[
\hat{\mu}_{i}=E_{i}\exp\left(\beta_{0}+\mathbf{Y}_{i}^{\left(\textrm{KILO}\right)}\boldsymbol{\hat{\beta}}^{\left(\textrm{KILO}\right)}+\mathbf{Y}_{i}^{\left(\textrm{ZONE}\right)}\boldsymbol{\hat{\beta}}^{\left(\textrm{ZONE}\right)}+\mathbf{Y}_{i}^{\left(\textrm{BONUS}\right)}\boldsymbol{\hat{\beta}}^{\left(\textrm{BONUS}\right)}+\mathbf{Y}_{i}^{\left(\textrm{MAKE}\right)}\boldsymbol{\hat{\beta}}^{\left(\textrm{MAKE}\right)}\right),
\]
for $i=1,2,\ldots,2182$. In Model I, these estimates immediately
generate an implied ``sample'' of the Poisson parameter, $\lambda$,
by setting $\hat{\lambda}_{i}=\hat{\mu}_{i}$. In Model II, they generate
an analogous ``sample'' of the Negative Binomial parameter, $p$,
by taking $\hat{p}_{i}=\tfrac{\hat{\mu}_{i}}{\hat{\mu}_{i}+\hat{r}}$.
Histograms of the $\hat{\lambda}_{i}$ and $\hat{p}_{i}$ are presented
in Figures 6 and 7, respectively.
\begin{singlespace}
\noindent \begin{center}
\medskip{}
\includegraphics[scale=0.7]{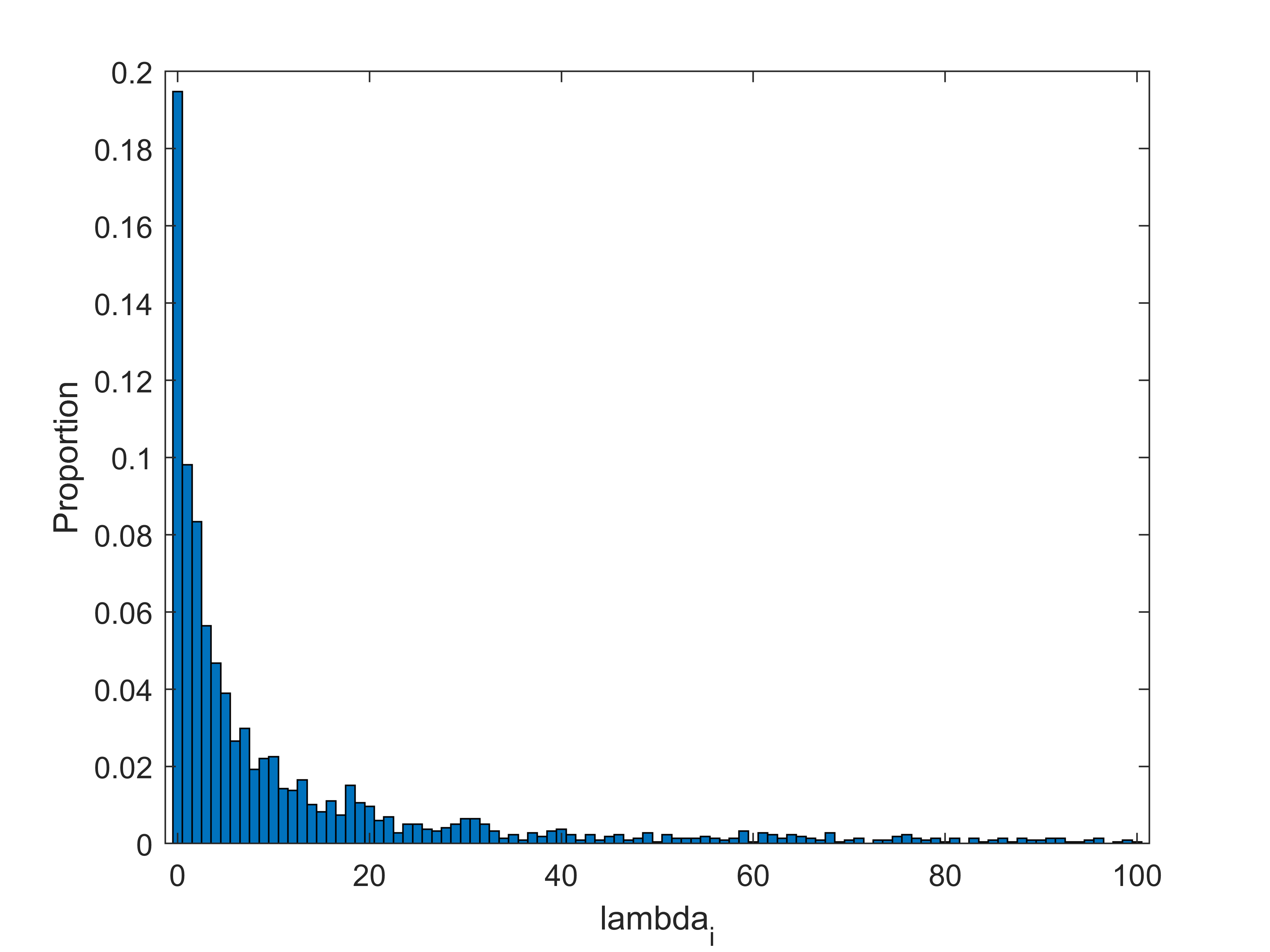}
\par\end{center}

\noindent \begin{center}
Figure 6. Histogram of Estimated $\lambda_{i}$ from Model I\medskip{}
\includegraphics[scale=0.7]{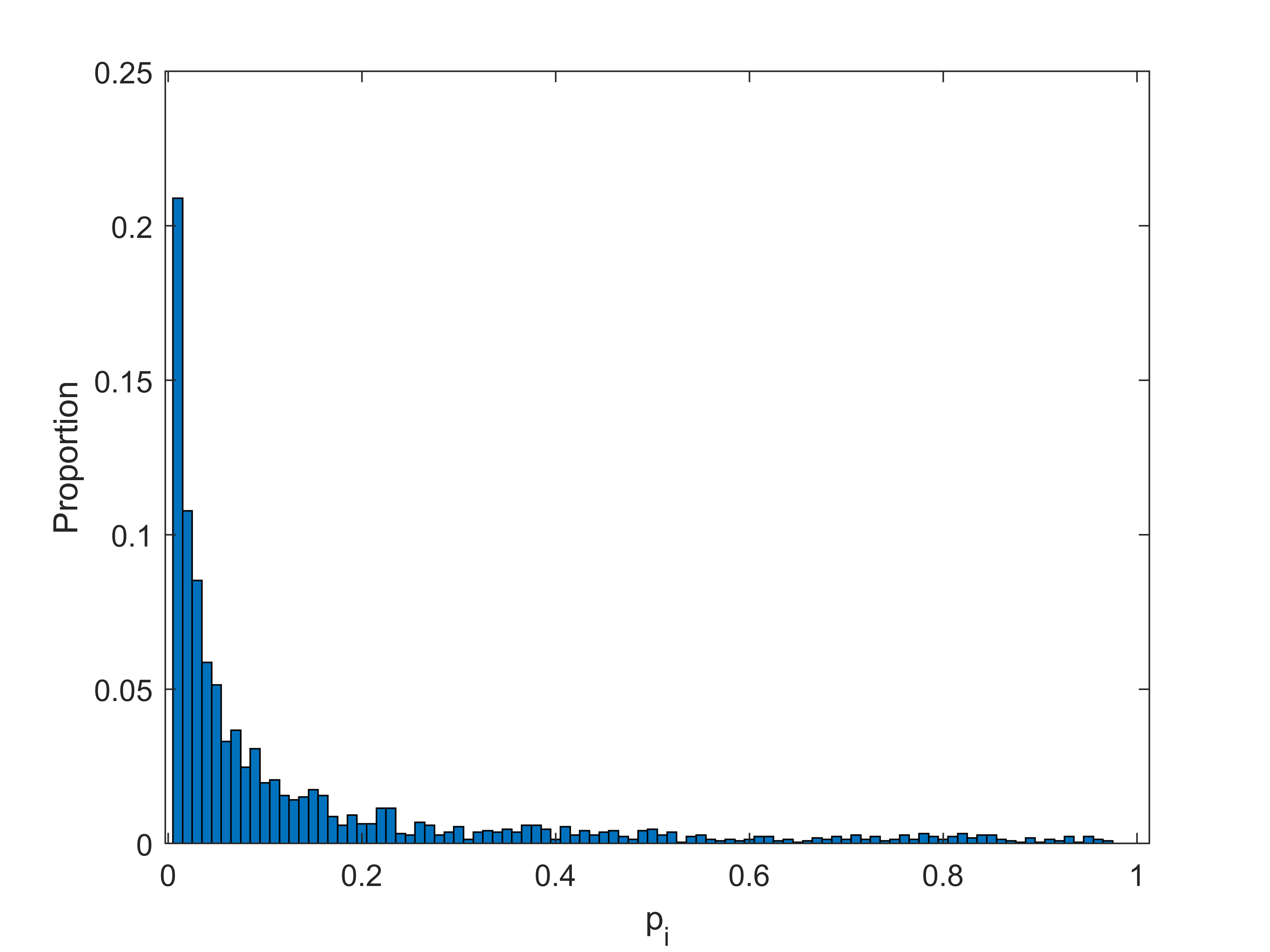}
\par\end{center}

\noindent \begin{center}
Figure 7. Histogram of Estimated $p_{i}$ from Model II\medskip{}
\par\end{center}
\end{singlespace}

\subsection{Testing Theoretical Mixing Distributions}

\noindent Theorem 4 of Subsection 3.3 allows one to calculate the
mixing PDFs for the Poisson parameter $\lambda$ associated with the
$\textrm{HGZY}\left(a,b,c,d\right)$, $\textrm{GZY}\left(a,b,c\right)$,
and $\textrm{Waring}\left(a,b\right)$ distributions. These theoretical
distributions then may be fit to the $\hat{\lambda}_{i}$ of Model
I, yielding the results presented in Tables 7 and 8. Since Model I
does not provide a good overall fit to the Swedish motor-vehicle data,
we would not expect the theoretical mixing distributions to fit the
$\hat{\lambda}_{i}$ very well; and that is indeed the case. Table
8 shows that the minimum-$\chi^{2}$ test rejects the null hypothesis
(that the $\hat{\lambda}_{i}$ come from the indicated mixing distribution)
at the 0.05 significance level for all three PDFs.\medskip{}
\begin{singlespace}
\noindent \begin{center}
Table 7. Maximum-Likelihood Estimation of $\lambda$ Mixing-Distribution
Parameters (Model I)
\par\end{center}

\noindent \begin{center}
\begin{tabular}{|c|c|c|c|c|c|}
\hline 
\textbf{\small{}Distribution} & \textbf{\small{}$\boldsymbol{\hat{a}}$ } & \textbf{\small{}$\boldsymbol{\hat{b}}$ } & \textbf{\small{}$\boldsymbol{\hat{c}}$ } & \textbf{\small{}$\boldsymbol{\hat{d}}$ } & \textbf{\small{}Log(Likelihood)}\tabularnewline
\hline 
{\small{}$f_{\lambda|a,b,c,d}\left(\lambda\right)$ for} & {\small{}0.0060} & {\small{}2.7456} & {\small{}740.1885} & {\small{}60.5348} & {\small{}-8529.05}\tabularnewline
{\small{}$\textrm{HGZY}\left(a,b,c,d\right)$} & {\small{}(0.0049){*}} & {\small{}(1.9963)} & {\small{}(640.6377)} & {\small{}(22.4654)} & \tabularnewline
\hline 
{\small{}$f_{\lambda|a,b,c}\left(\lambda\right)$ for} & {\small{}0.0650} & {\small{}0.9156} & {\small{}25.6792} &  & {\small{}-8525.61}\tabularnewline
{\small{}$\textrm{GZY}\left(a,b,c\right)$} & {\small{}(0.0195)} & {\small{}(0.0636)} & {\small{}(6.0218)} &  & \tabularnewline
\hline 
{\small{}$f_{\lambda|a,b}\left(\lambda\right)$ for} & {\small{}3.7199} & {\small{}0.7252} &  &  & {\small{}-8540.27}\tabularnewline
{\small{}$\textrm{Waring}\left(a,b\right)$} & {\small{}(0.2443)} & {\small{}(0.0272)} &  &  & \tabularnewline
\hline 
\end{tabular}
\par\end{center}
\end{singlespace}

\begin{singlespace}
\noindent \begin{flushleft}
{\small{}{*} Estimated standard errors of parameter estimators are
shown in parentheses.}\medskip{}
\par\end{flushleft}
\end{singlespace}

\begin{singlespace}
\noindent \begin{center}
Table 8. Goodness-of-Fit Measures for $\lambda$ Mixing Distributions
(Model I)
\par\end{center}

\noindent \begin{center}
\begin{tabular}{|c|c|c|c||c|c|}
\hline 
\textbf{\small{}Distribution} & \textbf{\small{}Min.-}{\small{}$\boldsymbol{\chi^{2}}$}\textbf{\small{}
Statistic} & \textbf{\small{}Degrees of Freedom} & \textbf{\small{}$\boldsymbol{p}$-Value} & \textbf{\small{}AIC} & \textbf{\small{}BIC}\tabularnewline
\hline 
{\small{}$f_{\lambda|a,b,c,d}\left(\lambda\right)$ for} & {\small{}44.14} & {\small{}28{*} - 4 - 1 = 23} & {\small{}0.0051} & {\small{}17,066.09} & {\small{}17,088.85}\tabularnewline
{\small{}$\textrm{HGZY}\left(a,b,c,d\right)$} &  &  &  &  & \tabularnewline
\hline 
{\small{}$f_{\lambda|a,b,c}\left(\lambda\right)$ for} & {\small{}41.69} & {\small{}31 - 3 - 1 = 27} & {\small{}0.0353} & {\small{}17,057.23} & {\small{}17,074.29}\tabularnewline
{\small{}$\textrm{GZY}\left(a,b,c\right)$} &  &  &  &  & \tabularnewline
\hline 
{\small{}$f_{\lambda|a,b}\left(\lambda\right)$ for} & {\small{}40.72} & {\small{}29 - 2 - 1 = 26} & {\small{}0.0331} & {\small{}17,084.54} & {\small{}17,095.91}\tabularnewline
{\small{}$\textrm{Waring}\left(a,b\right)$} &  &  &  &  & \tabularnewline
\hline 
\end{tabular}
\par\end{center}
\end{singlespace}

\begin{singlespace}
\noindent \begin{flushleft}
{\small{}{*} In computing the minimum-$\chi^{2}$ statistic for a
given mixing distribution, the cells consist of all intervals $\left(n,n+1\right]$,
where $n$ is a nonnegative integer such that $2182\times\Pr\left\{ \lambda\in\left(n,n+1\right]\right\} \geq10$,
plus 1 additional cell for all remaining integers $n$.\medskip{}
}
\par\end{flushleft}{\small \par}
\end{singlespace}

Theorem 2(2) of Subsection 3.2 permits the calculation of mixing PDFs
for the Negative Binomial parameter $p$, given the (fixed) estimated
value of $r$, $\hat{r}=110.5986$, for the same three members of
the $\mathcal{HG}$ family. Fitting these theoretical distributions
to the $p_{i}$ of Model II yields the results summarized in Tables
9 and 10. Since Model II provides a reasonable overall fit to the
motor-vehicle data, it is not surprising that the theoretical mixing
distributions also fit the $\hat{p}_{i}$ fairly well. In fact, Table
8 shows that the minimum-$\chi^{2}$ test does not reject the null
hypothesis (that the $\hat{p}_{i}$ come from the indicated mixing
distribution) at the 0.05 level for the PDFs associated with the $\textrm{GZY}\left(a,b,c\right)$
and $\textrm{Waring}\left(a,b\right)$ distributions.\medskip{}
\begin{singlespace}
\noindent \begin{center}
Table 9. Maximum-Likelihood Estimation of $p$ Mixing-Distribution
Parameters (Model II)
\par\end{center}

\noindent \begin{center}
\begin{tabular}{|c|c|c|c|c|c|}
\hline 
\textbf{\small{}Distribution} & \textbf{\small{}$\boldsymbol{\hat{a}}$ } & \textbf{\small{}$\boldsymbol{\hat{b}}$ } & \textbf{\small{}$\boldsymbol{\hat{c}}$ } & \textbf{\small{}$\boldsymbol{\hat{d}}$ } & \textbf{\small{}Log(Likelihood)}\tabularnewline
\hline 
{\small{}$f_{p|a,b,c,d}\left(p\right)$ for} & {\small{}0.0073} & {\small{}2.3350} & {\small{}592.7998} & {\small{}54.0534} & {\small{}2659.55}\tabularnewline
{\small{}$\textrm{HGZY}\left(a,b,c,d\right)$} & {\small{}(0.0062){*}} & {\small{}(1.4930)} & {\small{}(533.7293)} & {\small{}(22.7704)} & \tabularnewline
\hline 
{\small{}$f_{p|a,b,c}\left(p\right)$ for} & {\small{}0.0620} & {\small{}0.9187} & {\small{}26.3941} &  & {\small{}2664.30}\tabularnewline
{\small{}$\textrm{GZY}\left(a,b,c\right)$} & {\small{}(0.0183)} & {\small{}(0.0641)} & {\small{}(6.0980)} &  & \tabularnewline
\hline 
{\small{}$f_{p|a,b}\left(p\right)$ for} & {\small{}3.6690} & {\small{}0.7215} &  &  & {\small{}2648.56}\tabularnewline
{\small{}$\textrm{Waring}\left(a,b\right)$} & {\small{}(0.2407)} & {\small{}(0.0270)} &  &  & \tabularnewline
\hline 
\end{tabular}
\par\end{center}
\end{singlespace}

\begin{singlespace}
\noindent \begin{flushleft}
{\small{}{*} Estimated standard errors of parameter estimators are
shown in parentheses.\pagebreak{}}
\par\end{flushleft}{\small \par}
\end{singlespace}

\begin{singlespace}
\noindent \begin{center}
Table 10. Goodness-of-Fit Measures for $p$ Mixing Distributions (Model
II)
\par\end{center}

\noindent \begin{center}
\begin{tabular}{|c|c|c|c||c|c|}
\hline 
\textbf{\small{}Distribution} & \textbf{\small{}Min.-}{\small{}$\boldsymbol{\chi^{2}}$}\textbf{\small{}
Statistic} & \textbf{\small{}Degrees of Freedom} & \textbf{\small{}$\boldsymbol{p}$-Value} & \textbf{\small{}AIC} & \textbf{\small{}BIC}\tabularnewline
\hline 
{\small{}$f_{p|a,b,c,d}\left(p\right)$ for} & {\small{}39.80} & {\small{}28{*} - 4 - 1 = 23} & {\small{}0.0162} & {\small{}-5311.09} & {\small{}-5288.34}\tabularnewline
{\small{}$\textrm{HGZY}\left(a,b,c,d\right)$} &  &  &  &  & \tabularnewline
\hline 
{\small{}$f_{p|a,b,c}\left(p\right)$ for} & {\small{}36.34} & {\small{}31 - 3 - 1 = 27} & {\small{}0.1081} & {\small{}-5322.60} & {\small{}-5305.54}\tabularnewline
{\small{}$\textrm{GZY}\left(a,b,c\right)$} &  &  &  &  & \tabularnewline
\hline 
{\small{}$f_{p|a,b}\left(p\right)$ for} & {\small{}36.16} & {\small{}29 - 2 - 1 = 26} & {\small{}0.0887} & {\small{}-5293.13} & {\small{}-5281.75}\tabularnewline
{\small{}$\textrm{Waring}\left(a,b\right)$} &  &  &  &  & \tabularnewline
\hline 
\end{tabular}
\par\end{center}
\end{singlespace}

\begin{singlespace}
\noindent \begin{flushleft}
{\small{}{*} In computing the minimum-$\chi^{2}$ statistic for a
given mixing distribution, the cells consist of all intervals $\left(\tfrac{n}{n+\hat{r}},\tfrac{n+1}{n+1+\hat{r}}\right]$,
where $n$ is a nonnegative integer such that $2182\times\Pr\left\{ p\in\left(\tfrac{n}{n+\hat{r}},\tfrac{n+1}{n+1+\hat{r}}\right]\right\} \geq10$,
plus 1 additional cell for all remaining integers $n$.\medskip{}
}
\par\end{flushleft}{\small \par}
\end{singlespace}

The foregoing procedures demonstrate that an analysis of mixing distributions
can offer a useful means of validating conclusions based on GLM regression.
Although confirming the superiority of Model II to Model I may seem
somewhat redundant given the dispositive test results of Tables 5
and 6, it is important to keep in mind that model-selection decisions
are not always so clear-cut. This is especially true if one is choosing
from among numerous alternative models employing different sets of
risk classifications and other covariates.

Additionally, it should be noted that the analysis of mixing distributions
also can serve as an effective diagnostic tool. In the case at hand,
we see that the preferred model for fitting the motor-vehicle frequency
distribution is $\textrm{HGZY}\left(a,b,c,d\right)$, whereas the
preferred model for fitting the $\hat{p}_{i}$ of Model II is the
mixing distribution associated with $\textrm{GZY}\left(a,b,c\right)$.
This tells us that the GLM regression with Negative Binomial observations
discounts the impact of the $\mathcal{HG}$ family's $d$ parameter,
which plays a crucial role in governing the tail behavior of the frequency
distribution. Since increasing $d$ over the interval $\left[1,\infty\right)$
reduces the heaviness of the frequency tail, Model II effectively
overstates the impact of this tail. Although there are many potential
explanations for this deviation (e.g., collinearity among the explanatory
variables or the presence of statistical outliers), it is likely that
the assumption of a fixed Negative Binomial $r$ parameter plays a
major role. By applying exploratory GLM regressions to various subsets
of the motor-vehicle data, one can see that the $r$ parameter is
likely heterogeneous, requiring further modeling.

\section{Conclusions}

\noindent In the present article, we have investigated the class of
heavy-tailed frequency models formed by continuous mixtures of Negative
Binomial and Poisson random variables. We began by defining the concept
of a calibrative family of mixing distributions, each member of which
is identifiable from its associated Negative Binomial mixture, and
showed how to construct such families from only a single member. We
then introduced the heavy-tailed two-parameter ZY distribution as
a generalization of both the one-parameter Zeta and Yule distributions,
constructing calibrative families for both the new distribution and
the heavy-tailed two-parameter Waring distribution. Finally, we extended
both the ZY and Waring families to the four-parameter Hyper-Generalized
ZY family, which may be employed in conjunction with the concept of
calibrative families to analyze empirical loss-frequency data.

From both theoretical and applied perspectives, it is clear that the
assumption of a fixed Negative Binomial $r$ parameter is quite restrictive.
Given the general unavailability of calibrative families of joint
mixing distributions (of $p$ and $r$ simultaneously; see Subsection
3.2), we believe further research should focus on practical methods
for partitioning data into separate components within which the $r$
parameter is relatively homogeneous, while retaining the potential
benefits of analyzing mixing distributions.

As indicated at the outset, we hope that the present (and subsequent)
work encourages greater interest in the application of heavy-tailed
frequency models. However, we also would note that the study of calibrative
mixing distributions is equally useful for modeling continuous data,
such as loss severities. In the continuous case, the $\textrm{Gamma}\left(r,\lambda\right)$
distribution would constitute the natural analogue of the $\textrm{Negative Binomial}\left(r,p\right)$
distribution, with the $\textrm{Exponential}\left(\lambda\right)\equiv\textrm{Gamma}\left(r=1,\lambda\right)$
distribution taking over the central role of the $\textrm{Geometric}\left(p\right)\equiv\textrm{Negative Binomial}\left(r=1,p\right)$
distribution in constructing calibrative families.

\noindent \begin{center}
\textbf{\Large{}Appendix}
\par\end{center}{\Large \par}

\subsection*{{\small{}A.1 Proof of Theorem 1}}

\noindent {\small{}Since $E_{X\mid r,p}^{\left(\textrm{NB}\right)}\left[X^{\delta}\right]$
does not possess a convenient analytical form, we begin by considering
the case of $r\in\left\{ 1,2,3,\ldots\right\} $ and $\delta\in\left\{ 0,1,2,\ldots\right\} $,
so that
\[
E_{X\mid r,p}^{\left(\textrm{NB}\right)}\left[X^{\delta}\right]=E_{Z_{1},\ldots,Z_{r}\mid p}^{\left(\textrm{G}\right)}\left[\left(Z_{1}+Z_{2}+\cdots+Z_{r}\right)^{\delta}\right],
\]
where the $Z_{i}\sim\textrm{i.i.d. Geometric}\left(p\right)$. Then,
given that
\[
\prod_{j=1}^{k}E_{Z_{i}\mid p}^{\left(\textrm{G}\right)}\left[Z_{i}^{\delta_{j}}\right]\leq E_{Z_{i}\mid p}^{\left(\textrm{G}\right)}\left[Z_{i}^{\delta}\right]
\]
for any vector of positive integers $\left[\delta_{1},\delta_{2},\ldots,\delta_{k}\right]$
such that $\sum_{j=1}^{k}\delta_{j}=\delta$, we observe that
\[
r^{\delta}\left(E_{Z_{i}\mid p}^{\left(\textrm{G}\right)}\left[Z_{i}\right]\right)^{\delta}\leq E_{Z_{1},\ldots,Z_{r}\mid p}^{\left(\textrm{G}\right)}\left[\left(Z_{1}+Z_{2}+\cdots+Z_{r}\right)^{\delta}\right]\leq r^{\delta}E_{Z_{i}\mid p}^{\left(\textrm{G}\right)}\left[Z_{i}^{\delta}\right].
\]
For the $\textrm{Geometric}\left(p\right)$ distribution, it is known
that $E_{Z_{i}\mid p}^{\left(\textrm{G}\right)}\left[Z_{i}^{\delta}\right]=\tfrac{pA_{\delta}\left(p\right)}{\left(1-p\right)^{\delta}}$,
where $A_{\delta}\left(p\right)$ denotes the $\delta^{\textrm{th}}$-degree
Eulerian polynomial (which is of order $p^{\delta-1}$), so we can
rewrite the above inequalities as
\[
\left(\dfrac{rp}{1-p}\right)^{\delta}\leq E_{X\mid r,p}^{\left(\textrm{NB}\right)}\left[X^{\delta}\right]\leq\left(\dfrac{r}{1-p}\right)^{\delta}pA_{\delta}\left(p\right).\qquad\qquad\qquad\qquad\textrm{(A1)}
\]
}{\small \par}

{\small{}We now note that statement (a) is true if and only if $E_{X|r,\boldsymbol{\eta}}\left[X^{\delta}\right]=\underset{n\rightarrow\infty}{\lim}{\textstyle \sum_{x=0}^{n}}{\displaystyle x^{\delta}f_{X|r,\boldsymbol{\eta}}\left(x\right)}=\infty$
for some $\delta\in\left(0,\infty\right)$. This divergence is equivalent
to
\[
\underset{t\rightarrow1^{-}}{\lim}{\displaystyle \int_{0}^{t}E_{X\mid r,p}^{\left(\textrm{NB}\right)}\left[X^{\delta}\right]f_{p|\boldsymbol{\eta}}\left(p\right)dp}=\infty,
\]
which (by the right inequality of (A1)) implies
\[
\underset{t\rightarrow1^{-}}{\lim}{\displaystyle \int_{0}^{t}\left(\dfrac{r}{1-p}\right)^{\delta}pA_{\delta}\left(p\right)f_{p|\boldsymbol{\eta}}\left(p\right)dp}=\infty,\qquad\qquad\qquad\qquad\textrm{(A2)}
\]
and (by the left inequality of (A1)) is implied by
\[
\underset{t\rightarrow1^{-}}{\lim}{\displaystyle \int_{0}^{t}\left(\dfrac{rp}{1-p}\right)^{\delta}f_{p|\boldsymbol{\eta}}\left(p\right)dp}=\infty.\qquad\qquad\qquad\qquad\qquad\textrm{(A3)}
\]
}{\small \par}

{\small{}From statement (b), we know that, for some $\delta\in\left(0,\infty\right)$
and any $\varepsilon\in\left(0,L\right)$, there exists $p_{\varepsilon}\in\left(0,1\right)$
such that $p>p_{\varepsilon}\Longrightarrow f_{p|\boldsymbol{\eta}}\left(p\right)\left(1-p\right)^{-\delta+1}\geq L-\varepsilon>0$.
Thus,
\[
{\displaystyle \underset{t\rightarrow1^{-}}{\lim}\int_{0}^{t}\left(\dfrac{r}{1-p}\right)^{\delta}pA_{\delta}\left(p\right)f_{p|\boldsymbol{\eta}}\left(p\right)dp}\geq\int_{0}^{p_{\varepsilon}}\left(\dfrac{r}{1-p}\right)^{\delta}pA_{\delta}\left(p\right)f_{p|\boldsymbol{\eta}}\left(p\right)dp+\underset{t\rightarrow1^{-}}{\lim}\int_{p_{\varepsilon}}^{t}\dfrac{r^{\delta}pA_{\delta}\left(p\right)\left(L-\varepsilon\right)}{\left(1-p\right)}dp,
\]
where the second term on the right-hand side equals infinity, confirming
(A2). To show that condition (b) is implied by (A3)'s holding true
for some $\delta\in\left(0,\infty\right)$, assume the negation of
(b) (i.e., $\underset{p\rightarrow1^{-}}{\lim}f_{p|\boldsymbol{\eta}}\left(p\right)\left(1-p\right)^{-\delta+1}=0$
for all $\delta\in\left(0,\infty\right)$, which implies $\underset{p\rightarrow1^{-}}{\lim}f_{p|\boldsymbol{\eta}}\left(p\right)\left(1-p\right)^{-\delta+\kappa}=0$
for all $\delta\in\left(0,\infty\right)$ and any $\kappa\in\left(0,1\right)$).
This means that, for all $\delta\in\left(0,\infty\right)$ and any
$\varepsilon>0$, there exists $p_{\varepsilon}\in\left(0,1\right)$
such that $p>p_{\varepsilon}\Longrightarrow f_{p|\boldsymbol{\eta}}\left(p\right)\left(1-p\right)^{-\delta+\kappa}\leq\varepsilon$,
and so
\[
{\displaystyle \underset{t\rightarrow1^{-}}{\lim}\int_{0}^{t}\left(\dfrac{rp}{1-p}\right)^{\delta}f_{p|\boldsymbol{\eta}}\left(p\right)dp}\leq\int_{0}^{p_{\varepsilon}}\left(\dfrac{rp}{1-p}\right)^{\delta}f_{p|\boldsymbol{\eta}}\left(p\right)dp+{\displaystyle \underset{t\rightarrow1^{-}}{\lim}\int_{p_{\varepsilon}}^{t}\dfrac{\left(rp\right)^{\delta}\varepsilon}{\left(1-p\right)^{\kappa}}dp},
\]
where both terms on the right-hand side are finite, implying the negation
of (A3) for some $\delta\in\left(0,\infty\right)$.}{\small \par}

{\small{}To extend the above argument to all positive real-valued
$\delta$ and $r$, first note that
\[
\dfrac{\partial}{\partial\delta}E_{X\mid r,p}^{\left(\textrm{NB}\right)}\left[X^{\delta}\right]=\sum_{x=0}^{\infty}x^{\delta}\ln\left(x\right)\dfrac{\Gamma\left(x+r\right)}{\Gamma\left(r\right)\Gamma\left(x+1\right)}\left(1-p\right)^{r}p^{x}
\]
is positive for all $p\in\left(0,1\right)$ and $r\in\left(0,\infty\right)$,
and
\[
\dfrac{\partial}{\partial r}E_{X\mid r,p}^{\left(\textrm{NB}\right)}\left[X^{\delta}\right]=\sum_{x=0}^{\infty}x^{\delta}\dfrac{\Gamma\left(x+r\right)}{\Gamma\left(r\right)\Gamma\left(x+1\right)}\left(1-p\right)^{r}p^{x}\left\{ \dfrac{\Gamma^{\prime}\left(x+r\right)}{\Gamma\left(x+r\right)}-\dfrac{\Gamma^{\prime}\left(r\right)}{\Gamma\left(r\right)}+\ln\left(1-p\right)\right\} 
\]
is positive for all $p\in\left(p_{r},1\right)$ and $r\in\left(0,\infty\right)$,
for some sufficiently large $p_{r}$. We thus can extend the bounds
provided by (A1) to an arbitrary choice of $\delta\in\left(0,\infty\right)$
and $r\in\left[1,\infty\right)$ by writing
\[
\left(\dfrac{\left\lfloor r\right\rfloor p}{1-p}\right)^{\left\lfloor \delta\right\rfloor }\leq E_{X\mid r,p}^{\left(\textrm{NB}\right)}\left[X^{\delta}\right]\leq\left(\dfrac{\left\lceil r\right\rceil }{1-p}\right)^{\left\lceil \delta\right\rceil }pA_{\left\lceil \delta\right\rceil }\left(p\right)\qquad\qquad\qquad\qquad\textrm{(A4)}
\]
for $p\in\left(p_{\left\lfloor r\right\rfloor },1\right)$ (where
$\left\lfloor \cdot\right\rfloor $ and $\left\lceil \cdot\right\rceil $
denote the floor and ceiling functions, repsectively), and the equivalence
of statements (a) and (b) follows in the same way as before.}{\small \par}

{\small{}The remaining case of $r\in\left(0,1\right)$ is addressed
by employing the infinite-divisibility property of the Negative Binomial
distribution to express
\[
E_{Z_{1}\mid p}^{\left(\textrm{G}\right)}\left[Z_{1}^{\delta}\right]=E_{Z_{1}\mid p}^{\left(\textrm{G}\right)}\left[\left(X_{1}^{\left(u\right)}+X_{2}^{\left(u\right)}+\cdots+X_{m}^{\left(u\right)}\right)^{\delta}\right]
\]
and
\[
E_{Z_{2}\mid p}^{\left(\textrm{G}\right)}\left[Z_{2}^{\delta}\right]=E_{Z_{2}\mid p}^{\left(\textrm{G}\right)}\left[\left(X_{1}^{\left(\ell\right)}+X_{2}^{\left(\ell\right)}+\cdots+X_{m+1}^{\left(\ell\right)}\right)^{\delta}\right],
\]
where the $X_{i}^{\left(u\right)}\sim\textrm{i.i.d. Negative Binomial}\left(\tfrac{1}{m},p\right)$
and the $X_{i}^{\left(\ell\right)}\sim\textrm{i.i.d. Negative Binomial}\left(\tfrac{1}{m+1},p\right)$,
for $m=\left\lfloor \tfrac{1}{r}\right\rfloor $. By arguments similar
to those used earlier in the proof, it then can be shown that
\[
\dfrac{1}{m^{\delta}}E_{Z_{1}\mid p}^{\left(\textrm{G}\right)}\left[Z_{1}^{\delta}\right]\leq E_{X_{i}^{\left(u\right)}\mid p}^{\left(\textrm{NB}\right)}\left[\left(X_{i}^{\left(u\right)}\right)^{\delta}\right]\leq\dfrac{1}{m}E_{Z_{1}\mid p}^{\left(\textrm{G}\right)}\left[Z_{1}^{\delta}\right]
\]
\[
\Longleftrightarrow\dfrac{pA_{\delta}\left(p\right)}{m^{\delta}\left(1-p\right)^{\delta}}\leq E_{X_{i}^{\left(u\right)}\mid p}^{\left(\textrm{NB}\right)}\left[\left(X_{i}^{\left(u\right)}\right)^{\delta}\right]\leq\dfrac{pA_{\delta}\left(p\right)}{m\left(1-p\right)^{\delta}}
\]
and
\[
\dfrac{1}{\left(m+1\right)^{\delta}}E_{Z_{2}\mid p}^{\left(\textrm{G}\right)}\left[Z_{2}^{\delta}\right]\leq E_{X_{i}^{\left(\ell\right)}\mid p}^{\left(\textrm{NB}\right)}\left[\left(X_{i}^{\left(\ell\right)}\right)^{\delta}\right]\leq\dfrac{1}{\left(m+1\right)}E_{Z_{2}\mid p}^{\left(\textrm{G}\right)}\left[Z_{2}^{\delta}\right]
\]
\[
\Longleftrightarrow\dfrac{pA_{\delta}\left(p\right)}{\left(m+1\right)^{\delta}\left(1-p\right)^{\delta}}\leq E_{X_{i}^{\left(\ell\right)}\mid p}^{\left(\textrm{NB}\right)}\left[\left(X_{i}^{\left(\ell\right)}\right)^{\delta}\right]\leq\dfrac{pA_{\delta}\left(p\right)}{\left(m+1\right)\left(1-p\right)^{\delta}}.
\]
Since $E_{X_{i}^{\left(\ell\right)}\mid p}^{\left(\textrm{NB}\right)}\left[\left(X_{i}^{\left(\ell\right)}\right)^{\delta}\right]\leq E_{X\mid r,p}^{\left(\textrm{NB}\right)}\left[X^{\delta}\right]\leq E_{X_{i}^{\left(u\right)}\mid p}^{\left(\textrm{NB}\right)}\left[\left(X_{i}^{\left(u\right)}\right)^{\delta}\right]$
for all $p\in\left(p_{\tfrac{1}{m+1}},1\right)$, for some sufficiently
large $p_{\tfrac{1}{m+1}}$, these inequalities enable us to extend
the bounds of (A1) to an arbitrary choice of $\delta\in\left(0,\infty\right)$
and $r\in\left(0,1\right)$ via
\[
\dfrac{pA_{\left\lfloor \delta\right\rfloor }\left(p\right)}{\left(m+1\right)^{\left\lfloor \delta\right\rfloor }\left(1-p\right)^{\left\lfloor \delta\right\rfloor }}\leq E_{X\mid r,p}^{\left(\textrm{NB}\right)}\left[X^{\delta}\right]\leq\dfrac{pA_{\left\lceil \delta\right\rceil }\left(p\right)}{m\left(1-p\right)^{\left\lceil \delta\right\rceil }},\qquad\qquad\qquad\qquad\textrm{(A5)}
\]
and the desired result follows.}{\small \par}

{\small{}To demonstrate the equivalence of statements (b) and (c),
first let $R\left(p\right)=\tfrac{\ln\left(f_{p|\boldsymbol{\eta}}\left(p\right)\right)}{\ln\left(1-p\right)}$
and rewrite (b) as
\[
\underset{p\rightarrow1^{-}}{\lim}\left(1-p\right)^{R\left(p\right)-\delta+1}>0\textrm{ for some }\delta\in\left(0,\infty\right).
\]
Then
\[
\underset{p\rightarrow1^{-}}{\lim}\left(R\left(p\right)-\delta+1\right)\ln\left(1-p\right)>-\infty\textrm{ for some }\delta\in\left(0,\infty\right)
\]
\[
\Longleftrightarrow\underset{p\rightarrow1^{-}}{\lim}R\left(p\right)\leq\delta-1\textrm{ for some }\delta\in\left(0,\infty\right)
\]
\[
\Longleftrightarrow\underset{p\rightarrow1^{-}}{\lim}\dfrac{\ln\left(f_{p|\boldsymbol{\eta}}\left(p\right)\right)}{\ln\left(1-p\right)}<\infty.\:\blacksquare
\]
}{\small \par}

\subsection*{{\small{}A.2 Proof of Theorem 2}}

{\small{}For part (1), the uniqueness of $f_{p|\boldsymbol{\eta}}\left(p\right)$
follows immediately from the identifiability of Negative Binomial
mixtures with fixed $r$ (see Theorem 2.1 of Sapatinas, 1995). For
parts (2) and (3), we first derive the expression in (7), which holds
for all $r\in\left(0,\infty\right)$, and then show that (7) simplifies
to (6) for $r\in\left(1,\infty\right)$.}\footnote{{\small{}The expressions in parts (2) and (3) may be obtained as solutions
to Fredholm equations of the first type. However, we have found that
a few fortuitous rearrangements (see, e.g., equations (A6) and (A7))
obviate the need for such methods.}}{\small \par}

{\small{}First, note that $\int_{0}^{1}f_{X\mid r,p}^{\left(\textrm{NB}\right)}\left(x\right)f_{p|r,\boldsymbol{\eta}}\left(p\right)dp=f_{X|\boldsymbol{\eta}}\left(x\right)$
implies
\[
\int_{0}^{1}\dfrac{\Gamma\left(x+r\right)}{\Gamma\left(r\right)\Gamma\left(x+1\right)}\left(1-p\right)^{r}p^{x}f_{p|r,\boldsymbol{\eta}}\left(p\right)dp=\int_{0}^{1}\left(1-\omega\right)\omega^{x}f_{p|\boldsymbol{\eta}}\left(\omega\right)d\omega,
\]
or equivalently,
\[
\int_{0}^{1}\left(1-p\right)^{r}p^{x}f_{p|r,\boldsymbol{\eta}}\left(p\right)dp=\dfrac{\Gamma\left(r\right)\Gamma\left(x+1\right)}{\Gamma\left(x+r\right)}\int_{0}^{1}\left(1-\omega\right)\omega^{x}f_{p|\boldsymbol{\eta}}\left(\omega\right)d\omega
\]
\[
=\dfrac{\Gamma\left(r\right)\Gamma\left(x+1\right)}{\Gamma\left(x+r\right)}{\displaystyle \left[\dfrac{\left(x+1\right)}{\left(x+r\right)}\int_{0}^{1}\left(1-\omega\right)\omega^{x}f_{p|\boldsymbol{\eta}}\left(\omega\right)d\omega+\dfrac{\left(r-1\right)}{\left(x+r\right)}\int_{0}^{1}\left(1-\omega\right)\omega^{x}f_{p|\boldsymbol{\eta}}\left(\omega\right)d\omega\right]}
\]
\[
=\textrm{B}\left(x+1,r\right)\left[\int_{0}^{1}\omega^{x+1}f_{p|\boldsymbol{\eta}}\left(\omega\right)d\omega+\left(x+1\right)\int_{0}^{1}\omega^{x}f_{p|\boldsymbol{\eta}}\left(\omega\right)d\omega\right.
\]
\[
\left.-\left(x+2\right)\int_{0}^{1}\omega^{x+1}f_{p|\boldsymbol{\eta}}\left(\omega\right)d\omega+\left(r-1\right)\int_{0}^{1}\left(1-\omega\right)\omega^{x}f_{p|\boldsymbol{\eta}}\left(\omega\right)d\omega\right]\qquad\qquad\qquad\qquad\textrm{(A6)}
\]
\[
=\textrm{B}\left(x+1,r\right)\left[\int_{0}^{1}\omega^{x+1}f_{p|\boldsymbol{\eta}}\left(\omega\right)d\omega-\int_{0}^{1}\left(1-\omega\right)\omega^{x+1}f_{p|\boldsymbol{\eta}}^{\prime}\left(\omega\right)d\omega+\left(r-1\right)\int_{0}^{1}\left(1-\omega\right)\omega^{x}f_{p|\boldsymbol{\eta}}\left(\omega\right)d\omega\right]\qquad\textrm{(A7)}
\]
(where the middle integral in (A7) results from applying integration
by parts to each of the two middle integrals in (A6)). Since $\textrm{B}\left(x+1,r\right)={\textstyle \int_{0}^{1}t^{x}\left(1-t\right)^{r-1}dt}$,
it follows that
\[
\int_{0}^{1}\left(1-p\right)^{r}p^{x}f_{p|r,\boldsymbol{\eta}}\left(\omega\right)d\omega=\int_{0}^{1}t^{x}\left(1-t\right)^{r-1}dt\left[\int_{0}^{1}\omega^{x+1}f_{p|\boldsymbol{\eta}}\left(\omega\right)d\omega\right.
\]
\[
\left.-\int_{0}^{1}\left(1-\omega\right)\omega^{x+1}f_{p|\boldsymbol{\eta}}^{\prime}\left(\omega\right)d\omega+\left(r-1\right)\int_{0}^{1}\left(1-\omega\right)\omega^{x}f_{p|\boldsymbol{\eta}}\left(\omega\right)d\omega\right]
\]
\[
=\int_{0}^{1}\int_{0}^{1}\left(t\omega\right)^{x}\left(1-t\right)^{r-1}\omega f_{p|\boldsymbol{\eta}}\left(\omega\right)dtd\omega-\int_{0}^{1}\int_{0}^{1}\left(t\omega\right)^{x}\left(1-t\right)^{r-1}\omega\left(1-\omega\right)f_{p|\boldsymbol{\eta}}^{\prime}\left(\omega\right)dtd\omega
\]
\[
+\left(r-1\right)\int_{0}^{1}\int_{0}^{1}\left(t\omega\right)^{x}\left(1-t\right)^{r-1}\left(1-\omega\right)f_{p|\boldsymbol{\eta}}\left(\omega\right)dtd\omega.\qquad\qquad\qquad\qquad\qquad\textrm{(A8)}
\]
Substituting $p=\omega t$ into all three integrals of (A8) then yields
\[
\int_{0}^{1}\left(1-p\right)^{r}p^{x}f_{p|r,\boldsymbol{\eta}}\left(p\right)dp=\int_{0}^{1}\int_{0}^{\omega}p^{x}\left(1-\dfrac{p}{\omega}\right)^{r-1}f_{p|\boldsymbol{\eta}}\left(\omega\right)dpd\omega
\]
\[
-\int_{0}^{1}\int_{0}^{\omega}p^{x}\left(1-\dfrac{p}{\omega}\right)^{r-1}\left(1-\omega\right)f_{p|\boldsymbol{\eta}}^{\prime}\left(\omega\right)dpd\omega+\left(r-1\right)\int_{0}^{1}\int_{0}^{\omega}p^{x}\left(1-\dfrac{p}{\omega}\right)^{r-1}\dfrac{\left(1-\omega\right)}{\omega}f_{p|\boldsymbol{\eta}}\left(\omega\right)dpd\omega,
\]
\[
=\int_{0}^{1}\int_{0}^{\omega}p^{x}\dfrac{\left(\omega-p\right)^{r-1}}{\omega^{r-1}}f_{p|\boldsymbol{\eta}}\left(\omega\right)dpd\omega-\int_{0}^{1}\int_{0}^{\omega}p^{x}\dfrac{\left(\omega-p\right)^{r-1}\left(1-\omega\right)}{\omega^{r-1}}f_{p|\boldsymbol{\eta}}^{\prime}\left(\omega\right)dpd\omega
\]
\[
+\left(r-1\right)\int_{0}^{1}\int_{0}^{\omega}p^{x}\dfrac{\left(\omega-p\right)^{r-1}\left(1-\omega\right)}{\omega^{r}}f_{p|\boldsymbol{\eta}}\left(\omega\right)dpd\omega.\qquad\qquad\qquad\qquad\qquad\textrm{(A9)}
\]
Finally, by interchanging the order of integration and rearranging
components in (A9), we obtain
\[
\int_{0}^{1}\left(1-p\right)^{r}p^{x}f_{p|r,\boldsymbol{\eta}}\left(p\right)dp=\int_{0}^{1}\left(1-p\right)^{r}p^{x}\left[\dfrac{1}{\left(1-p\right)^{r}}\int_{p}^{1}\dfrac{\left(\omega-p\right)^{r-1}}{\omega^{r-1}}f_{p|\boldsymbol{\eta}}\left(\omega\right)d\omega\right]dp
\]
\[
-\int_{0}^{1}\left(1-p\right)^{r}p^{x}\left[\dfrac{1}{\left(1-p\right)^{r}}\int_{p}^{1}\dfrac{\left(\omega-p\right)^{r-1}\left(1-\omega\right)}{\omega^{r-1}}f_{p|\boldsymbol{\eta}}^{\prime}\left(\omega\right)\left(\omega\right)d\omega\right]dp
\]
\[
+\int_{0}^{1}\left(1-p\right)^{r}p^{x}\left[\dfrac{\left(r-1\right)}{\left(1-p\right)^{r}}\int_{p}^{1}\dfrac{\left(\omega-p\right)^{r-1}\left(1-\omega\right)}{\omega^{r}}f_{p|\boldsymbol{\eta}}\left(\omega\right)d\omega\right]dp
\]
\[
=\int_{0}^{1}\left(1-p\right)^{r}p^{x}\left\{ \dfrac{1}{\left(1-p\right)^{r}}\int_{p}^{1}\dfrac{\left(\omega-p\right)^{r-1}}{\omega^{r-1}}\left[1+\left(r-1\right)\dfrac{\left(1-\omega\right)}{\omega}\right]f_{p|\boldsymbol{\eta}}\left(\omega\right)d\omega\right\} dp
\]
\[
-\int_{0}^{1}\left(1-p\right)^{r}p^{x}\left[\dfrac{1}{\left(1-p\right)^{r}}\int_{p}^{1}\dfrac{\left(\omega-p\right)^{r-1}\left(1-\omega\right)}{\omega^{r-1}}f_{p|\boldsymbol{\eta}}^{\prime}\left(\omega\right)d\omega\right]dp,
\]
which implies (7).}{\small \par}

{\small{}If $r\in\left(1,\infty\right)$, then one can use integration
by parts to rewrite the second term on the right-hand side of (7)
as
\[
\dfrac{1}{\left(1-p\right)^{r}}\int_{p}^{1}\dfrac{\left(\omega-p\right)^{r-1}}{\omega^{r-1}}\left[\dfrac{p\left(r-1\right)\left(1-\omega\right)}{\left(\omega-p\right)\omega}-1\right]f_{p|\boldsymbol{\eta}}\left(\omega\right)d\omega,
\]
and substitute this expression to obtain (6). (Integration by parts
cannot be used for $r\in\left(0,1\right)$ because the resulting terms
diverge to infinity at $\omega=p$.)}{\small \par}

{\small{}The uniqueness of $f_{p|r>1,\boldsymbol{\eta}}\left(p\right)$
follows from the identifiability of Negative Binomial mixtures for
fixed $r$. If $f_{p|r<1,\boldsymbol{\eta}}\left(p\right)$ is a proper
PDF, then its uniqueness follows in the same way. The existence of
quasi-PDF solutions is shown by Theorem 3. $\blacksquare$}{\small \par}

\subsection*{{\small{}A.3 Proof of Theorem 3}}

\noindent {\small{}For part (1), consider the limit of (7) as $p\rightarrow0^{+}$:
\[
\underset{p\rightarrow0^{+}}{\lim}f_{p|r<1,\boldsymbol{\eta}}\left(p\right)=\underset{p\rightarrow0^{+}}{\lim}\dfrac{1}{\left(1-p\right)^{r}}\left[\left(2-r\right){\displaystyle \int_{p}^{1}}\left(\dfrac{\omega-p}{\omega}\right)^{r-1}f_{p|\boldsymbol{\eta}}\left(\omega\right)d\omega\right.
\]
\[
\left.+\left(r-1\right){\displaystyle \int_{p}^{1}}\left(\dfrac{\omega-p}{\omega}\right)^{r-1}\omega^{-1}f_{p|\boldsymbol{\eta}}\left(\omega\right)d\omega-{\displaystyle \int_{p}^{1}}\left(\dfrac{\omega-p}{\omega}\right)^{r-1}\left(1-\omega\right)f_{p|\boldsymbol{\eta}}^{\prime}\left(\omega\right)d\omega\right]
\]
\[
=2-r+\left(r-1\right)\underset{p\rightarrow0^{+}}{\lim}{\displaystyle \int_{p}^{1}}\omega^{-1}f_{p|\boldsymbol{\eta}}\left(\omega\right)d\omega-\left[\left.\left(1-\omega\right)f_{p|\boldsymbol{\eta}}\left(\omega\right)\right|_{0}^{1}+{\displaystyle \int_{0}^{1}}f_{p|\boldsymbol{\eta}}\left(\omega\right)d\omega\right]
\]
\[
=\underset{p\rightarrow0^{+}}{\lim}\left[f_{p|\boldsymbol{\eta}}\left(p\right)+\left(r-1\right){\displaystyle \int_{p}^{1}}\omega^{-1}f_{p|\boldsymbol{\eta}}\left(\omega\right)d\omega\right]-r+1.\qquad\qquad\qquad\qquad\textrm{(A10)}
\]
Clearly, $f_{p|r<1,\boldsymbol{\eta}}\left(p\right)$ is a quasi-PDF
with $\underset{p\rightarrow0^{+}}{\lim}f_{p|r<1,\boldsymbol{\eta}}\left(p\right)\leq\ell<0$
in some neighborhood of 0 if and only if (A10) is negative; or equivalently,
\[
\underset{p\rightarrow0^{+}}{\lim}{\displaystyle \int_{p}^{1}}\omega^{-1}f_{p|\boldsymbol{\eta}}\left(\omega\right)d\omega\left[\dfrac{f_{p|\boldsymbol{\eta}}\left(p\right)}{{\displaystyle \int_{p}^{1}}\omega^{-1}f_{p|\boldsymbol{\eta}}\left(\omega\right)d\omega}+r-1\right]<r-1.\qquad\qquad\qquad\textrm{(A11)}
\]
}{\small \par}

{\small{}If $\underset{p\rightarrow0^{+}}{\lim}f_{p|\boldsymbol{\eta}}\left(p\right)=0$,
then (A9) must hold because $\underset{p\rightarrow0^{+}}{\lim}{\textstyle \int_{p}^{1}}\omega^{-1}f_{p|\boldsymbol{\eta}}\left(\omega\right)d\omega=E_{\omega|\boldsymbol{\eta}}\left[\tfrac{1}{\omega}\right]>1$.
Moreover, if $0<\underset{p\rightarrow0^{+}}{\lim}f_{p|\boldsymbol{\eta}}\left(p\right)<\infty$,
then (A9) holds because $\underset{p\rightarrow0^{+}}{\lim}{\textstyle \int_{p}^{1}}\omega^{-1}f_{p|\boldsymbol{\eta}}\left(\omega\right)d\omega=\infty$.
Finally, if $\underset{p\rightarrow0^{+}}{\lim}f_{p|\boldsymbol{\eta}}\left(p\right)=\infty$
(implying $\underset{p\rightarrow0^{+}}{\lim}{\textstyle \int_{p}^{1}}\omega^{-1}f_{p|\boldsymbol{\eta}}\left(\omega\right)d\omega=\infty$
as well), then (A11) is satisfied if
\[
\underset{p\rightarrow0^{+}}{\lim}\left[\dfrac{f_{p|\boldsymbol{\eta}}\left(p\right)}{{\displaystyle \int_{p}^{1}}\omega^{-1}f_{p|\boldsymbol{\eta}}\left(\omega\right)d\omega}+r-1\right]<0
\]
\[
\Longleftrightarrow\underset{p\rightarrow0^{+}}{\lim}\dfrac{-pf_{p|\boldsymbol{\eta}}^{\prime}\left(p\right)}{f_{p|\boldsymbol{\eta}}\left(p\right)}<1-r,
\]
where the last inequality follows from L'Hôpital's rule. Condition
(A11) may or may not be true if $\underset{p\rightarrow0^{+}}{\lim}f_{p|\boldsymbol{\eta}}\left(p\right)=\infty$
and $\underset{p\rightarrow0^{+}}{\lim}\tfrac{-pf_{p|\boldsymbol{\eta}}^{\prime}\left(p\right)}{f_{p|\boldsymbol{\eta}}\left(p\right)}=1-r$,
because the expression on the left-hand side of (A11) becomes indeterminate.}{\small \par}

{\small{}For part (2), we first note that $\underset{p\rightarrow0^{+}}{\lim}f_{p|\boldsymbol{\eta}}\left(p\right)=\infty$
and $\underset{p\rightarrow0^{+}}{\lim}\tfrac{-pf_{p|\boldsymbol{\eta}}^{\prime}\left(p\right)}{f_{p|\boldsymbol{\eta}}\left(p\right)}\geq1-r$
(because (7) would be a quasi-PDF by (1)(a) otherwise), and rewrite
(7) as
\[
f_{p|r<1,\boldsymbol{\eta}}\left(p\right)=\dfrac{1}{\left(1-p\right)^{r}}{\displaystyle \int_{p}^{1}}\left(\dfrac{\omega-p}{\omega}\right)^{r-1}\left\{ 1-\left(1-\omega\right)\left[\dfrac{f_{p|\boldsymbol{\eta}}^{\prime}\left(\omega\right)}{f_{p|\boldsymbol{\eta}}\left(\omega\right)}-\dfrac{\left(r-1\right)}{\omega}\right]\right\} f_{p|\boldsymbol{\eta}}\left(\omega\right)d\omega.\qquad\textrm{(A12)}
\]
Then, since the inequality
\[
\dfrac{-pf_{p|\boldsymbol{\eta}}^{\prime}\left(p\right)}{f_{p|\boldsymbol{\eta}}\left(p\right)}>1-r-\dfrac{p}{1-p}
\]
is equivalent to
\[
1-\left(1-p\right)\left[\dfrac{f_{p|\boldsymbol{\eta}}^{\prime}\left(p\right)}{f_{p|\boldsymbol{\eta}}\left(p\right)}-\dfrac{\left(r-1\right)}{p}\right]>0,
\]
it follows that the expression inside the curly brackets \textendash{}
and therefore the entire integral in (A12) \textendash{} must be positive.
$\blacksquare$}{\small \par}

\subsection*{{\small{}A.4 Proof of Theorem 4}}

\noindent {\small{}Rewriting (8) with $r=1$ yields
\[
f_{X\mid r=1,p}^{\left(\textrm{NB}\right)}\left(x\right)=\int_{0}^{\infty}f_{X\mid\lambda}^{\left(\textrm{P}\right)}\left(x\right)f_{\lambda|r=1,\tfrac{1-p}{p}}^{\left(\Gamma\right)}\left(\lambda\right)d\lambda.
\]
Substituting the right-hand side of this equation into
\[
f_{X|\boldsymbol{\eta}}\left(x\right)=\int_{0}^{1}f_{X\mid r=1,p}^{\left(\textrm{NB}\right)}\left(x\right)f_{p|\boldsymbol{\eta}}\left(p\right)dp
\]
then gives
\[
f_{X|\boldsymbol{\eta}}\left(x\right)=\int_{0}^{1}\left[\int_{0}^{\infty}f_{X\mid\lambda}^{\left(\textrm{P}\right)}\left(x\right)f_{\lambda|r=1,\tfrac{1-p}{p}}^{\left(\Gamma\right)}\left(\lambda\right)d\lambda\right]f_{p|\boldsymbol{\eta}}\left(p\right)dp
\]
\[
=\int_{0}^{\infty}f_{X\mid\lambda}^{\left(\textrm{P}\right)}\left(x\right)\left[\int_{0}^{1}f_{\lambda|r=1,\tfrac{1-p}{p}}^{\left(\Gamma\right)}\left(\lambda\right)f_{p|\boldsymbol{\eta}}\left(p\right)dp\right]d\lambda
\]
\[
=\int_{0}^{\infty}f_{X\mid\lambda}^{\left(\textrm{P}\right)}\left(x\right)f_{\lambda|\boldsymbol{\eta}}\left(\lambda\right)d\lambda.\:\blacksquare
\]
}{\small \par}

\subsection*{{\small{}A.5 Proof of Theorem 5}}

\noindent {\small{}For part (A), consider
\[
f_{X|b,c>0}^{\left(\textrm{ZY}\right)}\left(x\right)={\displaystyle \int_{0}^{1}f_{X\mid r=1,p}^{\left(\textrm{NB}\right)}\left(x\right)f_{p|b,c>0}^{\left(\Sigma_{\textrm{B}}\right)}\left(p\right)dp}
\]
\[
={\displaystyle \int_{0}^{1}\left(1-p\right)p^{x}\dfrac{c\left(1-p^{c}\right)^{b}}{\Sigma_{\textrm{B}}\left(\dfrac{1}{c},\dfrac{1}{c},b\right)\left(1-p\right)}dp}
\]
\[
=\dfrac{c}{\Sigma_{\textrm{B}}\left(\dfrac{1}{c},\dfrac{1}{c},b\right)}{\displaystyle \int_{0}^{1}}\left(p^{c}\right)^{x/c}\left(1-p^{c}\right)^{b}dp.\qquad\qquad\qquad\qquad\qquad\textrm{(A13)}
\]
Substituting $q=p^{c}$ into the above integral allows us to rewrite
(A13) as
\[
\dfrac{1}{\Sigma_{\textrm{B}}\left(\dfrac{1}{c},\dfrac{1}{c},b\right)}{\displaystyle \int_{0}^{1}}q^{\left(x-c+1\right)/c}\left(1-q\right)^{b}dq
\]
\[
=\dfrac{1}{\Sigma_{\textrm{B}}\left(\dfrac{1}{c},\dfrac{1}{c},b\right)}\textrm{B}\left(\dfrac{x+1}{c},b+1\right).
\]
}{\small \par}

\noindent {\small{}For part (B), note that
\[
f_{X|b,c\rightarrow0}^{\left(\textrm{ZY}\right)}\left(x\right)={\displaystyle \int_{0}^{1}f_{X\mid r=1,p}^{\left(\textrm{NB}\right)}\left(x\right)f_{p|b,c\rightarrow0}^{\left(\Sigma_{\textrm{B}}\right)}\left(p\right)dp}
\]
\[
={\displaystyle \int_{0}^{1}\left(1-p\right)p^{x}\dfrac{\left(-\ln\left(p\right)\right)^{b}}{\zeta\left(b+1\right)\Gamma\left(b+1\right)\left(1-p\right)}dp}
\]
\[
=\dfrac{1}{\zeta\left(b+1\right)\Gamma\left(b+1\right)}{\displaystyle \int_{0}^{1}}p^{x}\left(-\ln\left(p\right)\right)^{b}dp.\qquad\qquad\qquad\qquad\qquad\qquad\textrm{(A14)}
\]
Then, using the substitution $t=-\ln\left(p\right)$ in the above
integral, (A14) can be rewritten as
\[
\dfrac{1}{\zeta\left(b+1\right)\Gamma\left(b+1\right)}{\displaystyle \int_{\infty}^{0}}\left(e^{-t}\right)^{x}t^{b}\left(-e^{-t}\right)dt
\]
\[
=\dfrac{1}{\zeta\left(b+1\right)\Gamma\left(b+1\right)}{\displaystyle \int_{0}^{\infty}}\left(e^{-t}\right)^{x+1}t^{b}dt
\]
\[
=\dfrac{\left(x+1\right)^{-\left(b+1\right)}}{\zeta\left(b+1\right)}{\displaystyle \int_{0}^{\infty}}\dfrac{\left(x+1\right)^{b+1}t^{b}e^{-\left(x+1\right)t}}{\Gamma\left(b+1\right)}dt
\]
\[
=\dfrac{\left(x+1\right)^{-\left(b+1\right)}}{\zeta\left(b+1\right)}.\:\blacksquare
\]
}{\small \par}

\subsection*{{\small{}A.6 Proof of Corollary 1}}

\noindent {\small{}In parts (A)(1) and (A)(2), the functional forms
follow immediately by replacing $f_{p|\boldsymbol{\eta}}\left(p\right)$
by (12) in parts (2) and (3), respectively, of Theorem 2. In parts
(B)(1) and (B)(2), the functional forms similarly follow by replacing
$f_{p|\boldsymbol{\eta}}\left(p\right)$ by (13) in parts (2) and
(3), respectively, of Theorem 2.}{\small \par}

{\small{}The fact that $f_{p|r<1,b,c>0}\left(p\right)\leq\ell<0$
for $p$ in some neighborhood of 0 follows from condition (1)(a) of
Theorem 3, since $\underset{p\rightarrow0^{+}}{\lim}f_{p|b,c>0}^{\left(\Sigma_{\textrm{B}}\right)}\left(p\right)=\tfrac{c}{\Sigma_{\textrm{B}}\left(\tfrac{1}{c},\tfrac{1}{c},b\right)}\in\left(0,\infty\right)$.
Likewise, $f_{p|r<1,b,c\rightarrow0}\left(p\right)\leq\ell<0$ in
a neighborhood of 0 because $\underset{p\rightarrow0^{+}}{\lim}f_{p|b,c\rightarrow0}^{\left(\Sigma_{\textrm{B}}\right)}\left(p\right)=\infty$
and $\underset{p\rightarrow0^{+}}{\lim}\tfrac{-pf_{p|b,c\rightarrow0}^{\left(\Sigma_{\textrm{B}}\right)\prime}\left(p\right)}{f_{p|b,c\rightarrow0}^{\left(\Sigma_{\textrm{B}}\right)}\left(p\right)}=0<1-r$,
thus satisfying the same condition of Theorem 3. $\blacksquare$}{\small \par}

\subsection*{{\small{}A.7 Proof of Corollary 2}}

\noindent {\small{}For part (A), replacing $f_{p|\boldsymbol{\eta}}\left(p\right)$
by (12) in Theorem 4 gives
\[
f_{\lambda|b,c>0}\left(\lambda\right)={\displaystyle \int_{0}^{1}}\left(\dfrac{1-p}{p}\right)\exp\left(-\left(\dfrac{1-p}{p}\right)\lambda\right)\dfrac{c\left(1-p^{c}\right)^{b}}{\Sigma_{\textrm{B}}\left(\dfrac{1}{c},\dfrac{1}{c},b\right)\left(1-p\right)}dp
\]
\[
=\dfrac{c}{\Sigma_{\textrm{B}}\left(\dfrac{1}{c},\dfrac{1}{c},b\right)}{\displaystyle \int_{0}^{1}}\dfrac{1}{p}\left(1-p^{c}\right)^{b}\exp\left(-\left(\dfrac{1-p}{p}\right)\lambda\right)dp.
\]
Substituting $y=\tfrac{1-p}{p}$ into this integral then yields the
expression on the right-hand side of (14).}{\small \par}

{\small{}For part (B), replacing $f_{p|\boldsymbol{\eta}}\left(p\right)$
by (13) in Theorem 4 gives
\[
f_{\lambda|b,c\rightarrow0}\left(\lambda\right)={\displaystyle \int_{0}^{1}}\left(\dfrac{1-p}{p}\right)\exp\left(-\left(\dfrac{1-p}{p}\right)\lambda\right)\dfrac{\left(-\ln\left(p\right)\right)^{b}}{\zeta\left(b+1\right)\Gamma\left(b+1\right)\left(1-p\right)}dp
\]
\[
=\dfrac{1}{\zeta\left(b+1\right)\Gamma\left(b+1\right)}{\displaystyle \int_{0}^{1}}\dfrac{1}{p}\left(-\ln\left(p\right)\right)^{b}\exp\left(-\left(\dfrac{1-p}{p}\right)\lambda\right)dp.
\]
Substituting $y=\tfrac{1-p}{p}$ into this integral then yields the
right-hand side of (15). $\blacksquare$}{\small \par}

\subsection*{{\small{}A.8 Proof of Corollary 3}}

\noindent {\small{}In part (1), the functional form follows immediately
by replacing $f_{p|\boldsymbol{\eta}}\left(p\right)$ by
\[
f_{p|a,b}^{\left(\textrm{B}\right)}\left(p\right)=\dfrac{1}{\textrm{B}\left(a,b\right)}p^{a-1}\left(1-p\right)^{b-1}\qquad\qquad\qquad\qquad\qquad\textrm{(A15)}
\]
in part (2) of Theorem 2. In parts (2) and (3), the functional form
similarly follows by replacing $f_{p|\boldsymbol{\eta}}\left(p\right)$
by (A15) in part (3) of Theorem 2.}{\small \par}

{\small{}To show that $f_{p|r<1,a,b}\left(p\right)$ is the unique
mixing PDF when $r\geq a$, first note that $\underset{p\rightarrow0^{+}}{\lim}f_{p|a<1,b}^{\left(\textrm{B}\right)}\left(p\right)=\infty$
and $\underset{p\rightarrow0^{+}}{\lim}\tfrac{-pf_{p|a<1,b}^{\left(\textrm{B}\right)\prime}\left(p\right)}{f_{p|a<1,b}^{\left(\textrm{B}\right)}\left(p\right)}=1-a\geq1-r$.
Then, since $\tfrac{-pf_{p|a<1,b}^{\left(\textrm{B}\right)\prime}\left(p\right)}{f_{p|a<1,b}^{\left(\textrm{B}\right)}\left(p\right)}=\tfrac{bp}{1-p}+2-a-\tfrac{1}{1-p}>1-r-\tfrac{p}{1-p}$
for all $p\in\left(0,1\right)$, the result follows from condition
(2)(a) of Theorem 3.}{\small \par}

{\small{}To show that $f_{p|r<1,a,b}\left(p\right)\leq\ell<0$ for
all $p$ in some neighborhood of 0 when $r<a$, we must consider three
cases: $a>1$, $a=1$, and $a<1$. For $a>1$ and $a=1$, the result
follows from condition (1)(a) of Theorem 3, since $\underset{p\rightarrow0^{+}}{\lim}f_{p|a>1,b>0}^{\left(\textrm{B}\right)}\left(p\right)=0$
and $\underset{p\rightarrow0^{+}}{\lim}f_{p|a=1,b>0}^{\left(\textrm{B}\right)}\left(p\right)=b\in\left(0,\infty\right)$,
respectively. For $a<1$, the same condition of Theorem 3 applies,
since $\underset{p\rightarrow0^{+}}{\lim}f_{p|a<1,b}^{\left(\textrm{B}\right)}\left(p\right)=\infty$
and $\underset{p\rightarrow0^{+}}{\lim}\tfrac{-pf_{p|a<1,b}^{\left(\textrm{B}\right)\prime}\left(p\right)}{f_{p|a<1,b}^{\left(\textrm{B}\right)}\left(p\right)}=1-a<1-r$.
$\blacksquare$}{\small \par}

\subsection*{{\small{}A.9 Proof of Corollary 4}}

\noindent {\small{}Replacing $f_{p|\boldsymbol{\eta}}\left(p\right)$
by (A15) in Theorem (4) gives
\[
f_{\lambda|a,b}\left(\lambda\right)={\displaystyle \int_{0}^{1}}\left(\dfrac{1-p}{p}\right)\exp\left(-\left(\dfrac{1-p}{p}\right)\lambda\right)\dfrac{p^{a-1}\left(1-p\right)^{b-1}}{\textrm{B}\left(a,b\right)}dp
\]
\[
=\dfrac{1}{\textrm{B}\left(a,b\right)}{\displaystyle \int_{0}^{1}}p^{a-1}\left(1-p\right)^{b-1}\exp\left(-\left(\dfrac{1-p}{p}\right)\lambda\right)dp.
\]
Substituting $y=\tfrac{1-p}{p}$ into this integral then yields the
expression on the right-hand side of (16). $\blacksquare$}{\small \par}
\end{document}